\PassOptionsToPackage{table,dvipsnames}{xcolor}
\documentclass{article} 
\usepackage{iclr2027_conference,times}

\usepackage{microtype}
\usepackage{amsfonts}
\usepackage{amsmath}
\usepackage{amsthm}
\usepackage{xcolor}
\usepackage{graphicx}     
\usepackage{booktabs}     
\usepackage{tabularx}
\usepackage{longtable}
\usepackage{xltabular}
\usepackage{makecell}
\usepackage{multirow}
\usepackage{wrapfig}
\usepackage{float}
\usepackage{subcaption}
\usepackage{caption}
\usepackage{enumitem}     
\usepackage{pifont}       
\usepackage{url}
\usepackage{adjustbox}
\usepackage[most]{tcolorbox} 
\usepackage{fontawesome5}    
\usepackage{algpseudocode}
\usepackage[linesnumbered,lined,boxed,commentsnumbered,ruled,longend]{algorithm2e}
\usepackage[T1]{fontenc}

\usepackage{tikz}
\usepackage[edges]{forest}
\usetikzlibrary{
  arrows.meta,
  positioning,
  calc,
  shapes.symbols,
  shapes.geometric,
  shapes.misc
}

\usepackage{hyperref}
\usepackage[noabbrev,nameinlink]{cleveref}
\hypersetup{
    colorlinks=true,
    linkcolor=black,
    citecolor=cyan,
    filecolor=magenta,
    urlcolor=magenta,
    }

\theoremstyle{plain}

\theoremstyle{definition}
\newtheorem{definition}{Definition}[section]

\theoremstyle{remark}

\crefname{definition}{Definition}{Definitions}
\Crefname{definition}{Definition}{Definitions}

\definecolor{themecolor}{HTML}{37D2A6}        
\definecolor{themecolor_light}{HTML}{9BE9D3}
\definecolor{themecolor_lighter}{HTML}{CDF4E9}

\newcommand\blfootnote[1]{%
  \begingroup
  \renewcommand\thefootnote{}\footnote{#1}%
  \addtocounter{footnote}{-1}%
  \endgroup
}

\newcommand{\ghlink}[1]{\faIcon{github}\,\href{#1}{GitHub}}
\newcommand{\weblink}[1]{\faIcon{globe}\,\href{#1}{Website}}

\title{AgentSWE: Can Coding Agents Build\\
the Agent You Actually Want?}

\newcommand{\coderepo}{\url{https://github.com/VectorSpaceLab/AgentSWE}}

\author{Jiahao Wang$^{1,2,\ast}$ \quad Hongjin Qian$^{1,\ast}$ \quad
        Yuyang Hu$^{1,3}$ \quad Jiajun Zhang$^{2}$ \quad Zheng Liu$^{1,\dagger}$ \\
$^{1}$Beijing Academy of Artificial Intelligence \\
$^{2}$Institute of Automation, Chinese Academy of Sciences \\
$^{3}$Renmin University of China \\
Code: \coderepo
}

\iclrfinalcopy

\begin{document}

\maketitle
\lhead{Preprint}

\begin{abstract}
\blfootnote{$^{\ast}$Equal contribution. \quad $^{\dagger}$Corresponding author.}%
 As the production of software passes into the hands of software itself,
  coding agents are increasingly commissioned to build agents fitted to their users' own workflows. This delegation carries a
quiet assumption, that the delivered agent will faithfully serve the requirement
it was built for, rather than merely appear to. No existing benchmark tests
that assumption.  We introduce \emph{AgentSWE}, a benchmark for
\emph{agent software engineering} spanning three stages of the agent
development lifecycle: \textbf{Creation} of a complete agent from a natural-language
requirement, \textbf{Editing} of production agent codebases, and
\textbf{Optimization} toward a measurable target. Each of its 25
tasks is framed as a commission rather than an exam: the builder iterates
against a handful of illustrative cases, while acceptance exercises the
frozen submission on held-out cases designed to span the full requirement,
grading behavior against requirement-derived rubrics under hard pass/fail
gates or, for Optimization, against the source benchmark's own objective.
Evaluating a range of coding agents across models and harnesses, we find
that none comes close to delivering what a requirement asks for:
the gap between what a builder demonstrates and what it delivers is the
norm, not the corner case. An accompanying empirical analysis of the
  failure trajectories identifies recurring failure modes, among them
  overfitting to the visible cases, hollow runtimes behind code that looks
  complete, and unsupported claims of completion, each a signal by which
  coding agents can learn to build the agents they are asked for.
\end{abstract}

\section{Introduction}
\label{sec:intro}

Coding agents such as Claude Code~\citep{anthropic2026claudecode}
and Codex~\citep{openai2026codexcli} now carry substantial engineering
workloads~\citep{chen2021evaluating,yang2024sweagent,wang2024openhands}, and
many of their users are not engineers: they describe what they need and
delegate the rest~\citep{hitzig2026agentic}.  Increasingly, what they
delegate is the building of an agent, a program that wraps a model, tools,
and state around one recurring job so that the job runs on a fixed runtime,
at a fixed cost, without a coding agent attending each run.
Delegation of this kind rests on trust: that a coding agent, handed a
requirement in natural language, will deliver the agent its user intends
rather than one that merely appears to comply.  Whether that trust is
warranted is what this paper measures.

\begin{figure}[t]
    \centering
    \includegraphics[width=\linewidth]{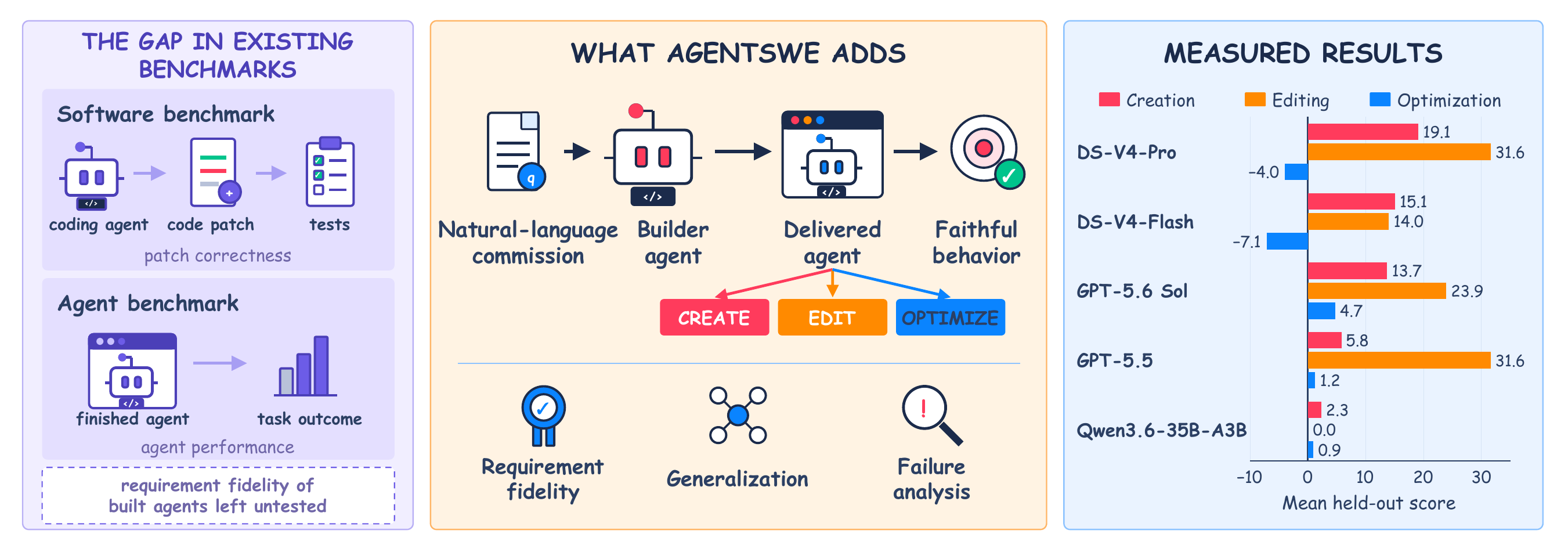}
    \caption{
        AgentSWE at a glance.
        (a) Existing benchmarks grade the patch a coding agent writes, or the
        performance of an agent someone else built; neither asks whether a
        coding agent delivers the agent a requirement describes.
        (b) AgentSWE commissions an agent from a natural-language requirement
        and grades what the delivered agent does on held-out cases, over
        creation, editing and optimization.
        (c) Mean held-out Result ($S$ for Optimization) of five builders in
        the Codex harness (\Cref{tab:main}).
    }
    \label{fig:overview}
\end{figure}

Agent development spans three stages. It begins with
\emph{creation}, where a natural-language requirement  must become a
working system; continues through \emph{editing}, where new capabilities join
a living codebase without unsettling what already works; and matures into
\emph{optimization}, where the agent is pushed toward an explicit target, be
it a public leaderboard or a production workload. Handing this lifecycle to
coding agents invites failures subtler than broken code:
an agent that passes visible examples yet misses the intent behind
them~\citep{liu2023evalplus,zhao2026specbench,lodkaew2026deceive}; a tool registered but
never invoked~\citep{soni2026toolfailbench,fan2026agentprocessbench}; a goal
that drifts over a long
horizon~\citep{arike2025goaldrift,menon2026inherited}; a report citing
evidence no source contains~\citep{onweller2026cited,rao2026reference}. Such
failures are not hypothetical; they are the present condition of agent-built
agents~\citep{raj2026model,naik2026guardianagentbench}. The questions
are how far coding agents compromise the requirements they are handed, on
which capabilities, and for what reasons.

These questions outrun the instruments we have.  SWE-bench and its
descendants grade the repair of conventional software against a reference
patch or test suite~\citep{swebench,mlebench}; NL2Repo-Bench commissions
whole repositories from requirement documents~\citep{ding2025nl2repobench};
Paper2Code and PaperBench rebuild research implementations checked against
the papers they reproduce~\citep{seo2025paper2code,starace2025paperbench}.
Each judges ordinary code against an answer known in advance.  A
commissioned agent wraps a stochastic policy around tools, answers to a
requirement that has no reference solution, and keeps changing after
delivery.  Benchmarks that do evaluate agents, from AgentBench
onward~\citep{liu2023agentbench}, examine a finished agent and ask how well
it performs; none asks whether a coding agent could have built it.
Work on agents that design or improve other
agents~\citep{hu2025adas,mishra2026aiebench} comes closest, but it measures
how far a search procedure or a meta-agent can raise a benchmark score, not
whether a general coding agent delivers the agent a requirement describes
(\Cref{sec:related}).
\textbf{Between judging built software and judging finished agents lies a
step that current suites leave unmeasured: building the agent that a
requirement asks for.}

We introduce \emph{AgentSWE}, a benchmark that turns faithful delegation
from anecdote into measurement (\Cref{fig:overview}) through three
commitments. The first is to measure the whole lifecycle, through
twenty-five tasks spanning all three stages. Ten \emph{Creation} tasks
commission a complete agent from a natural-language requirement, over
workloads from GUI automation to formal theorem proving, each scoped against
a mature open-source system; ten \emph{Editing} tasks ask for new capabilities in the
pinned codebase of a production agent such as OpenHands or Aider,
capabilities that cut through crash recovery, permission fencing, and
transactional state; and five \emph{Optimization} tasks set a working agent against a
measurable target. The second commitment is that evaluation should mirror the
delegation it studies. Each task is framed as a commission: \textbf{the
builder sees only what a customer could provide, while acceptance also tests
the stated clauses its examples leave out.} The builder develops against the
requirement and a few illustrative cases; held-out cases then test the frozen
delivery in isolation. The third commitment follows from the absence of a
prescribed reference implementation: every Creation and Editing execution
is graded against a rubric derived from the requirement itself, and hard
gates assign zero to invalid executions and inapplicable patches.

We put this instrument to work on five models as builders in the Codex
harness, together with further pairings of harness and model. The verdict is
sobering. The strongest builders average 19.1 on Creation and 31.6 on
Editing, and in Optimization no builder closes more than 5\,\% of the
remaining gap to the target on average, and only 13 of the 25 runs score
above their starter. Creation runs that end development at 90 or above
deliver agents averaging 9.9 on held-out cases.
\textbf{The gap between what a builder demonstrates and what it delivers is not the
corner case of agent software engineering; it is the norm.} The builders'
trajectories reveal recurring failures. Builders overfit to the visible development cases,
and the delivery fails on the clauses those cases leave out; they ship
hollow runtimes, code that reads as complete but does no work when run; and
both builders and the agents they deliver claim completions that their own
records contradict. Each of these modes names a habit that the next
generation of coding agents can be built and trained to avoid. More
development cases, a prescribed design, an instruction to test, and a
longer case budget shift the scores but leave held-out Result low under
every condition.

In sum, this paper makes three contributions.  First, we formalize agent software
engineering as a measurable problem, in which a builder must satisfy a
requirement it can probe only through limited development feedback.
Second, we deliver AgentSWE, twenty-five commissions whose acceptance
executes the delivered agent on held-out cases and grades each run with a
requirement-derived rubric or a native objective, and we evaluate various harness–model pairings against
it.  Third, we contribute an empirical analysis of the failure trajectories, which identifies recurring failure modes of coding
agents and turns each into a direction for the next generation of coding
agents.

\section{Related Work}
\label{sec:related}

\paragraph{Agents that build and optimize agents.}
Research on automated agent development has expanded both the scope of what can be designed and the stage at which adaptation occurs. Early approaches generate task-specific roles, tools, and collaboration patterns from task descriptions or seed agents~\citep{chen2024autoagents,yuan2025evoagent,tang2025autoagent}. Subsequent work formulates agent development as optimization, first over textual components such as instructions and demonstrations~\citep{yang2024large,khattab2024dspy,agrawal2026gepa}, and then over broader design spaces containing agent modules, communication structures, and executable workflows~\citep{zhuge2024gptswarm,shang2025agentsquare,zhang2025aflow,zhang2025maas,zhou2026mass}. This paradigm has further expanded to searching over complete code-defined agents and multi-agent scaffolds~\citep{hu2025adas,rosser2025agentbreeder}.

A related line relaxes the assumption that the improvement procedure itself remains fixed. Initial systems support self-referential modification of an improver or agent logic~\citep{zelikman2023stop,yin2024godel}, while later approaches expose complete implementations, harnesses, and meta-level improvement mechanisms to revision~\citep{robeyns2025sica,zhang2025dgm,wang2026hgm,zhang2026hyperagents,lee2026metaharness,hebbar2026sia}. Arbor~\citep{jin2026arbor} carries hypotheses and evidence across long-horizon artifact optimization, whereas JIT-Agent~\citep{zhang2026jitagent} synthesizes task-conditioned executable harnesses at inference time. Together, these developments establish agent construction and optimization as a capability in its own right. AgentSWE makes this capability the object of evaluation, measuring whether general-purpose coding agents can create, modify, and optimize agents from heterogeneous user requirements.

\paragraph{Benchmarks for agent construction.}
Most agent benchmarks hold the agent implementation fixed and evaluate its performance within a given environment. Recent benchmarks move beyond task completion by asking agents to construct or iteratively improve more complex artifacts. WorFBench~\citep{qiao2025worfbench} evaluates generated workflow structures, while benchmarks for machine-learning experimentation and research engineering evaluate improvement of executable research artifacts~\citep{huang2024mlagentbench,nathani2025mlgym,wijk2024rebench}. AI4AI-Bench~\citep{chi2026ai4aibench} extends this setting to the redesign of model-training algorithms. Other evaluations isolate specific improvement mechanisms: SE-Bench~\citep{yuan2026sebench} studies knowledge internalization, whereas OPT-BENCH~\citep{li2026optbench} measures feedback-driven solution refinement. These benchmarks evaluate workflows, research artifacts, or optimized solutions rather than deployable agents.

Closest to our setting, AIE-Bench~\citep{mishra2026aiebench} evaluates whether a meta-agent can improve a pinned target-agent scaffold under controlled mutation boundaries and trial budgets. AgentSWE extends this setting to the full agent-development lifecycle: creating agents from natural-language commissions, modifying existing agent codebases without regression and optimizing working agents toward measurable objectives. By freezing submissions and evaluating their behavior on held-out cases, AgentSWE measures whether coding agents can deliver the agents users actually request.
\section{Problem Formulation}
\label{sec:formulation}

A commission prescribes no reference implementation, so we define the
requirement, the delivered agent, the builder, and the fidelity of a
delivery through observable behavior alone.

\subsection{Preliminaries}
\label{sec:formulation-prelim}

\paragraph{Requirements.}
A commission begins with a natural-language requirement
$q=(g,\mathcal{I},\mathcal{C},\mathcal{R})$, which every AgentSWE task
publishes in four parts: the goal $g$, which states what the agent is for
and what counts as serving its user; the interface contract $\mathcal{I}$,
which fixes how the agent is launched and which artifacts it must deliver;
the constraints $\mathcal{C}$, which bound what may be edited, installed,
or invoked; and the resources $\mathcal{R}$, such as data, repositories,
and access to models and search.  A \emph{case} $x$ is one concrete
instance of the commission.  The requirement comes with a small set $D$ of
public development cases, while the cases that matter, the ones the
delivered agent will face in use, remain unwritten; we write
$\mathcal{X}_q$ for the space of cases the requirement implies.

\paragraph{Agents.}
The deliverable is itself an agent $a=(h,m)$: a \emph{harness} $h$, the orchestration code that plans, routes, and retries~\citep{raj2026model,lee2026metaharness}, around a
\emph{backbone model} $m$ reached through a fixed gateway.  Executing $a$
on a case $x$ in environment $E$ yields observable outcomes
$y \sim a(x;E)$.  Because $m$ samples, the same delivery may satisfy a
requirement on one execution and betray it on the next, so acceptance
executes the agent and grades what it does.

\paragraph{Builders.}
A coding agent, or \emph{builder}, is an agent $\beta$ whose actions are
software engineering: reading a requirement, exploring and editing a
workspace, running what it wrote, and deciding what to try next
\citep{yang2024sweagent,wang2024openhands}.  Handed a
commission, the builder produces an artifact
$\hat{a} \sim \beta(q, a_0, D; B)$, where $a_0$ is the initial artifact,
possibly empty, $D$ the development cases it may probe freely, and $B$ a
budget on time and compute.

\subsection{The Agent-for-Agent Problem}
\label{sec:formulation-problem}

The customer cares about how faithfully $\hat a$ serves the requirement on
cases it has not seen.  Let $\mu_q(x, y) \in [0,1]$ be the graded
satisfaction of requirement $q$ on case $x$ with outcome $y$, and let
$V(\hat a) \in \{0,1\}$ be a validity gate on the non-negotiables of
$\mathcal{I}$ and $\mathcal{C}$: the agent launches, its deliverables parse,
its changes respect the authorized surface.  The \emph{fidelity} of a
delivered agent is
\begin{equation}
F\bigl(\hat a;\, q\bigr)
\;=\;
V(\hat a)\cdot
\mathop{\mathbb{E}}_{x \sim \mathcal{X}_q}\;
\mathop{\mathbb{E}}_{y \sim \hat a(x;E)}
\bigl[\, \mu_q(x, y) \,\bigr].
\label{eq:fidelity}
\end{equation}

\begin{definition}[Agent software engineering]
\label{def:ase}
Given a requirement $q$, an initial artifact $a_0$, development cases
$D$, and a budget $B$, a builder $\beta$ must deliver an agent
$\hat a \sim \beta(q, a_0, D; B)$ maximizing fidelity
$F(\hat a; q)$.
\end{definition}

We estimate fidelity on a held-out set $\bar D \subset \mathcal{X}_q$,
disjoint from $D$ and designed to exercise the promises in $g$, graded by a
rubric that realizes $\mu_q$.  Writing $F_D$ and $F_{\bar D}$ for the
estimates on the development and held-out cases, the gap between
demonstrated and delivered fidelity, $F_D - F_{\bar D}$, measures how much better the delivered agent
does on the cases the builder saw than on the cases it did not.

\subsection{Task Families}
\label{sec:formulation-families}

The three lifecycle stages are instances of \Cref{def:ase} that differ in
the initial artifact $a_0$ and in the shape of $\mu_q$.

\paragraph{Creation ($a_0 = \emptyset$).}
The builder starts from an empty workspace and must deliver a complete
agent under the interface contract $\mathcal{I}$; fidelity is estimated on
held-out cases as in \Cref{eq:fidelity}.

\paragraph{Editing ($a_0 = $ a production system).}
The initial artifact is the pinned codebase of a working agent product, and
the requirement commissions new capabilities.  The constraints
$\mathcal{C}$ confine the builder's changes to an authorized region, and
satisfaction has two parts,
\begin{equation}
\mu_q \;=\; \mu_{\mathrm{new}} \,\wedge\, \mu_{\mathrm{reg}},
\label{eq:edit}
\end{equation}
the commissioned behavior $\mu_{\mathrm{new}}$ and the preservation
$\mu_{\mathrm{reg}}$ of everything the product already did, checked by its
public suite and published contracts \citep{swebench}.

\paragraph{Optimization ($a_0 = $ a working agent).}
The initial artifact already satisfies the interface, and the requirement
names a measurable objective $J$ on a 0--100 scale, whose ceiling
$J^{\star}=100$ is the target.  Delivery is judged by the fraction of the remaining gap to the target that
the builder closes
\citep[cf.][]{wijk2024rebench},
\begin{equation}
\gamma(\hat a)
\;=\;
\min\!\left\{1,\;
\frac{J(\hat a) - J(a_0)}{\,J^{\star} - J(a_0)\,}\right\},
\label{eq:opt}
\end{equation}
on held-out instances of the objective.

\section{The AgentSWE Benchmark}
\label{sec:benchmark}

\begin{figure}[t]
    \centering
    \includegraphics[width=0.94\linewidth]{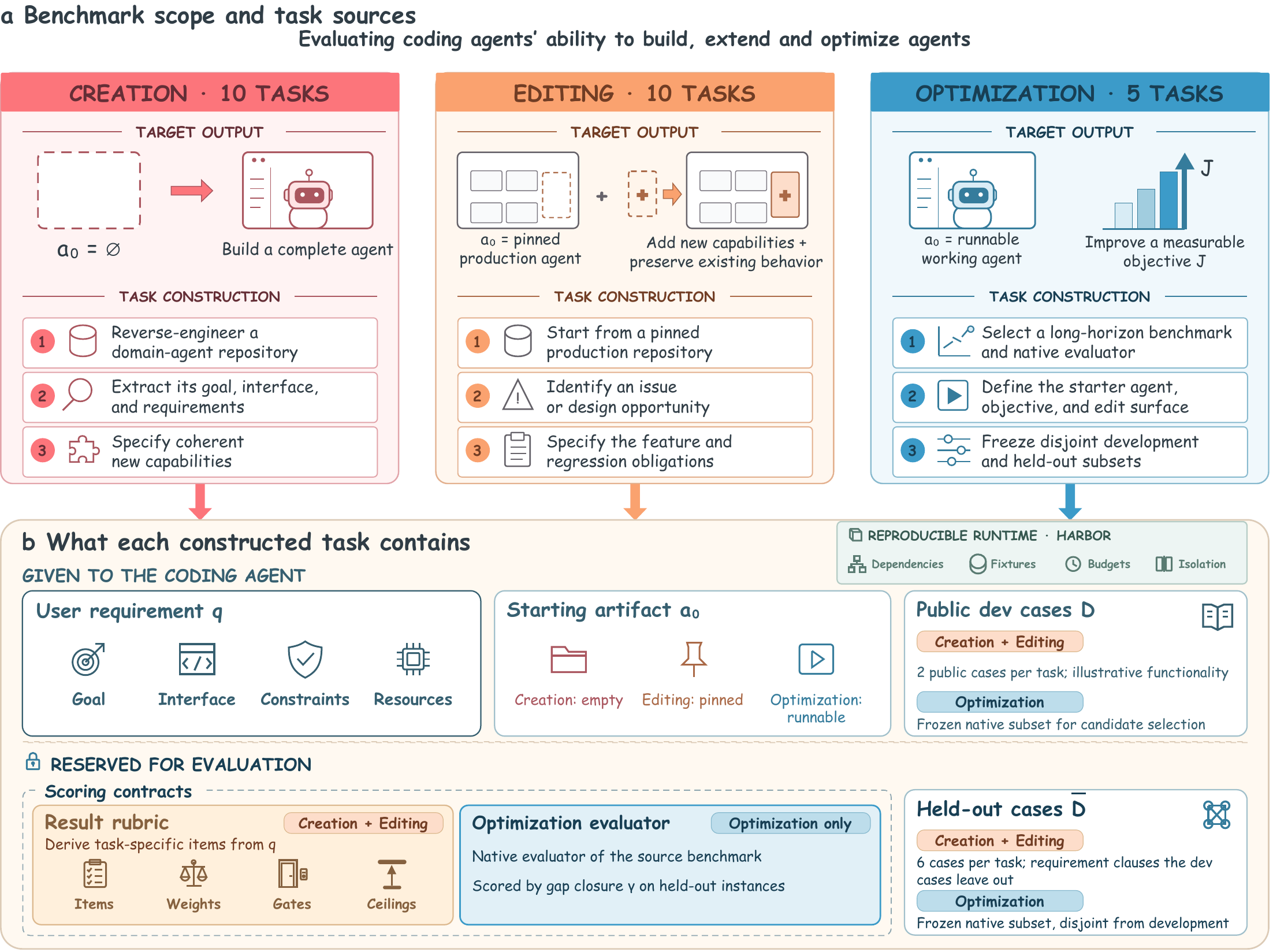}
    \caption{
        AgentSWE task families and benchmark construction.
        (a) Task objectives and construction processes.
        (b) Builder-visible inputs, evaluator-owned scoring contracts
        and held-out cases, supported by Harbor environments.
    }
    \label{fig:benchmark-construction}
\end{figure}

AgentSWE turns \Cref{def:ase} into 25 commissions: ten Creation, ten
Editing, and five Optimization tasks.  Each task hands a coding agent, the
builder, a requirement $q$, an initial artifact $a_0$, and development cases
$D$, and grades the agent it delivers on held-out cases.
\Cref{fig:benchmark-construction,fig:evaluation-protocol} show how tasks are
built and evaluated; \Cref{app:tasks} walks through one task per family.

\subsection{Task Construction}
\label{sec:benchmark-sourcing}

Each family fixes the requirement $q$ and the initial artifact $a_0$ in its
own way (\Cref{fig:benchmark-construction}a).  In every family, GPT-5.6 Sol~\citep{openai2026gpt56} drafts the task materials from the
analyzed source system, and the authors verify and refine each draft.
Creation tasks come from open-source domain agents on GitHub, in domains
from desktop automation and scientific PDF translation to formal theorem
proving.  Each repository is reverse-engineered into a self-contained goal
and a set of functional requirements, extended with capabilities that
broaden its purpose and raise the compositional difficulty, and published
as input Markdown files that state the goal, interface contract,
constraints, and resources of $q$; the initial artifact is empty.

Editing tasks extend a pinned repository snapshot $a_0$ of five
general-purpose coding agents and five domain agents, among them Claude
Code~\citep{anthropic2026claudecode}, Aider~\citep{gauthier2026aider}, OpenHands~\citep{wang2024openhands}, and
OpenClaw~\citep{openclawcontributors2026openclaw}.  The commissioned capabilities come
from design brainstorming or from unresolved issues in the repository; we
keep proposals with a consequential user need, a bounded implementation
surface, and behavior that execution can verify, and write each as a
feature specification with an acceptance boundary and the regression
obligations that $a_0$ already carries.

Optimization tasks build on BrowseComp~\citep{wei2025browsecomp}, $\tau^3$~\citep{yao2024taubench,barres2025tau2bench}, TerminalBench~\citep{merrill2026terminalbench}, PinchBench~\citep{kiloai2026pinchbench}, and OSWorld~\citep{xie2024osworld} and keep their native task definitions.  Each ships a 
starter agent as $a_0$ (\Cref{tab:task-sources}), and the requirement names
the objective, the files the builder may edit, and the runtime resources it
may not change.

\subsection{Task Package}
\label{sec:benchmark-package}

Each task separates what the builder sees, namely $q$, $a_0$, and the
development cases $D$, from what evaluation reserves
(\Cref{fig:benchmark-construction}b).  Every Creation and Editing task
publishes two development cases, which demonstrate the entry point, the
delivery contract, and a small part of the required functionality;
Optimization draws a development split from the instance pool of its source
benchmark.

Evaluation reserves the held-out cases and the scoring contract.  Each
Creation and Editing task holds out six cases.  Each case exercises one or
more requirement clauses, among them clauses and capability interactions
that the requirement states but the development cases leave out, and
together the six cover diverse
inputs, state, and data.  Optimization holds out a disjoint split of the
same pool.  In total the suite has 150 development and 365 held-out
cases.  The scoring contract of a Creation or Editing task is a task-specific
\emph{Result} rubric whose weighted items derive from the functional
requirements, with gates and ceilings for violations of the interface
contract and the constraints (\Cref{app:rubrics}); Optimization keeps the
native evaluator of its source benchmark.  Held-out inputs, evaluator state,
scoring contracts, credentials, and model routing never enter the builder's
workspace, and the builder receives feedback only on its development cases.

\subsection{Evaluation Protocol}
\label{sec:protocol}

\begin{figure}[t]
    \centering
    \includegraphics[width=\linewidth]{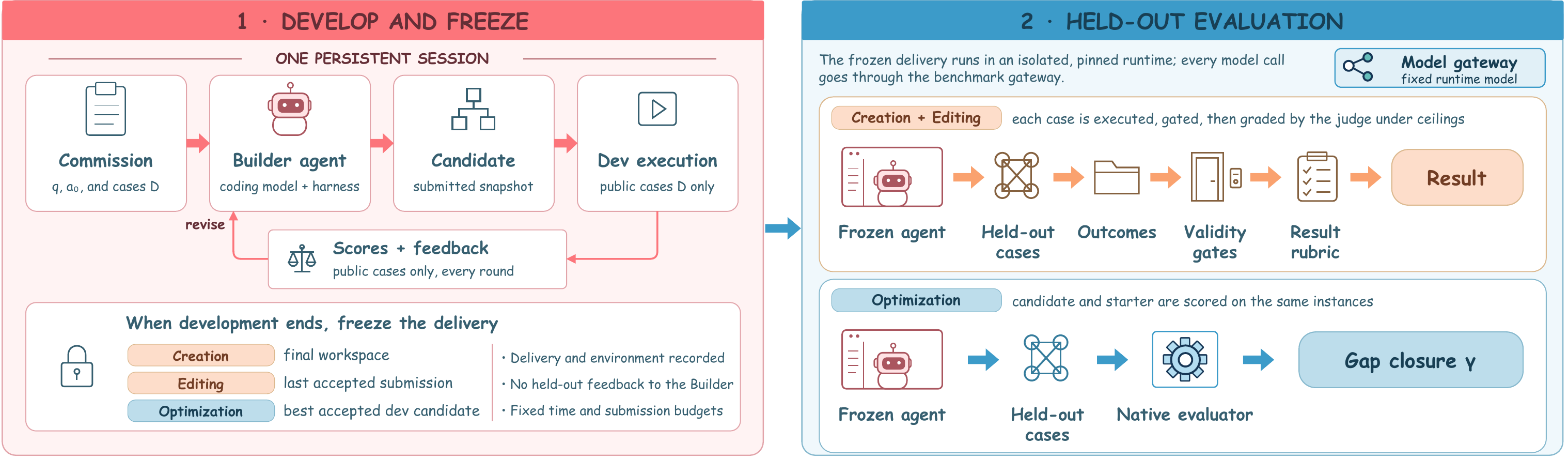}
    \caption{
        AgentSWE evaluation workflow: development with public feedback,
        delivery selection and freezing, and held-out evaluation using
        task-specific Result rubrics or native benchmark objectives.
    }
    \label{fig:evaluation-protocol}
\end{figure}

\paragraph{Development and freezing.}
Harbor~\citep{harborteam2026harbor}, a
container framework for agent evaluation, runs the builder in one persistent
session (\Cref{fig:evaluation-protocol}, left).  The builder works in a
private writable workspace and may submit a snapshot of its delivery at any
time.  Each snapshot runs on the development cases in isolated jobs, and the
scores and summarized feedback return to the same session, within fixed time
and submission budgets (\Cref{app:setup}).  Development ends when the builder
exits or exhausts its submissions or its time budget.  Creation then freezes
the workspace; Editing freezes the last accepted submission, so a
builder whose submissions were never accepted delivers nothing; Optimization freezes
the accepted snapshot with the highest development objective.  Freezing
records the delivery and its software environment, and the held-out
evaluation verifies that neither has changed.

\paragraph{Held-out execution.}
Each held-out case runs the frozen delivery through its fixed entry point in
an isolated, pinned runtime (\Cref{fig:evaluation-protocol}, right;
\Cref{app:infra}).  Every model call the delivered agent makes goes through a
benchmark-owned gateway to one runtime model, shared by all deliveries in a
task family, so that the score reflects what the builder contributes:
prompts, tool orchestration, context and state management, and output
validation.

\paragraph{Scoring.}
For Creation and Editing, deterministic gates first assign zero to an execution that does not
launch, exceeds its time budget, or produces no parseable deliverable.  A frontier reasoning model at its highest effort then applies the Result
rubric~\citep{zheng2023judging,arora2025healthbench} (GPT-5.6 Sol for Creation, DeepSeek-V4.1-Flash~\citep{deepseekai2026deepseekv41flash} for
Editing; \Cref{tab:setup}).  The judge grades what the agent produced on one
concrete case, its artifacts, execution trace, and resulting state, rather
than the quality of its code, and ceilings bind the score to the violations
the evaluator records.  The task score is the mean partial-credit score
over its held-out cases on a 0--100 scale, and family scores are macro-averages over tasks.
Optimization instead scores the selected candidate and the starter with the
native evaluator on the same held-out instances, a signal independent of
the Result judge, and reports the gap closure of \Cref{eq:opt}.

\section{Experiments}
\label{sec:experiments}

\Cref{sec:exp-main} measures
held-out performance, \Cref{sec:exp-analysis} analyzes the trajectories of
these runs and the failure modes they reveal, and \Cref{sec:exp-discussion}
examines how held-out Result and these failures respond when the conditions
of the build change or the build is repeated.

\subsection{Setup}
\label{sec:exp-setup}

The main comparison evaluates five builders that pair the Codex harness with
GPT-5.6 Sol, GPT-5.5~\citep{openai2026gpt55}, DeepSeek-V4-Pro, DeepSeek-V4-Flash~\citep{deepseekai2026deepseekv4}, and
Qwen3.6-35B-A3B~\citep{qwen36_35b_a3b}; we refer to each builder by its model. A \emph{run} is one
builder developing one task, from the commission to the frozen delivery.
Creation and Editing report the mean Result over the six held-out cases of
each task. Optimization reports gap closure
$S=100\gamma$ (\Cref{eq:opt}), computed from the change
$\Delta=J(\hat a)-J(a_0)$ in the native objective on the held-out split;
$S$ is negative when the candidate scores below its starter.
\Cref{app:setup} gives the settings for each task family;
\Cref{app:limitations} discusses the limitations of the evaluation.

\subsection{Held-out Performance}
\label{sec:exp-main}

\begin{table}[t]
\centering
\caption{Held-out performance across the three lifecycle stages. Creation
and Editing report mean Result (0--100); Optimization reports gap closure
$S=100\gamma$. \Cref{app:opt-full} gives per-run Optimization results. DS
denotes DeepSeek.}
\label{tab:main}
\small
\setlength{\tabcolsep}{4pt}
\begin{tabular}{@{}lrrrrr@{}}
\toprule
\textbf{Task} & \textbf{GPT-5.6 Sol} & \textbf{GPT-5.5} & \textbf{DS-V4-Pro} & \textbf{DS-V4-Flash} & \textbf{Qwen3.6} \\
\midrule
\multicolumn{6}{@{}l}{\textit{Creation}} \\
Authorized vulnerability validation & 0.0 & 0.0 & 14.0 & 14.0 & 0.0 \\
Database analytics & 3.0 & 9.8 & 14.3 & 0.0 & 0.0 \\
Desktop GUI automation & 18.0 & 3.2 & 17.5 & 12.0 & 0.0 \\
Document-to-editable PPTX & 0.0 & 9.8 & 18.7 & 16.8 & 21.7 \\
Evidence-grounded document QA & 3.3 & 12.2 & 0.0 & 3.5 & 0.0 \\
Formal theorem proving & 66.7 & 0.0 & 16.5 & 63.0 & 0.0 \\
Repository bug repair & 27.7 & 0.0 & 60.0 & 0.0 & 0.0 \\
Schema-guided web extraction & 0.0 & 0.0 & 6.7 & 0.0 & 0.0 \\
Scientific PDF translation & 18.3 & 21.8 & 25.3 & 26.0 & 0.0 \\
Web research report & 0.0 & 0.8 & 18.0 & 15.3 & 1.7 \\
\quad Mean & 13.7 & 5.8 & 19.1 & 15.1 & 2.3 \\
\midrule
\multicolumn{6}{@{}l}{\textit{Editing}} \\
Claude policy provenance & 24.0 & 43.7 & 28.3 & 0.0 & 0.0 \\
Aider worktree transaction & 21.5 & 20.0 & 47.3 & 0.0 & 0.0 \\
OpenHands effect recovery & 4.0 & 20.5 & 24.3 & 0.0 & 0.0 \\
OpenClaw channel handoff & 38.5 & 12.3 & 19.5 & 0.0 & 0.0 \\
Codex execution residual & 11.7 & 70.0 & 35.0 & 0.0 & 0.0 \\
AI-Scientist reproducibility gate & 48.5 & 67.3 & 61.7 & 54.3 & 0.0 \\
DeepCode claim traceability & 26.5 & 41.3 & 32.7 & 47.0 & 0.0 \\
DeepTutor adaptive remediation & 54.2 & 14.3 & 31.0 & 25.7 & 0.0 \\
Dyad acceptance-driven workflow & 9.8 & 18.0 & 22.5 & 0.0 & 0.0 \\
OpenWiki change-impact documentation & 0.0 & 8.5 & 14.0 & 12.8 & 0.0 \\
\quad Mean & 23.9 & 31.6 & 31.6 & 14.0 & 0.0 \\
\midrule
\multicolumn{6}{@{}l}{\textit{Optimization}} \\
BrowseComp & 32.3 & 18.4 & $-$1.0 & 34.1 & 8.5 \\
$\tau^3$ & 4.3 & 0.0 & 15.8 & $-$50.0 & $-$23.8 \\
TerminalBench & $-$10.0 & 7.7 & 0.0 & 2.8 & 7.7 \\
PinchBench & $-$14.9 & $-$11.1 & $-$38.7 & $-$34.4 & 12.1 \\
OSWorld & 12.0 & $-$9.1 & 4.0 & 12.0 & 0.0 \\
\quad Mean & 4.7 & 1.2 & $-$4.0 & $-$7.1 & 0.9 \\
\bottomrule
\end{tabular}
\end{table}

\paragraph{Creation success is rare and task-specific.}
The 50 Creation runs in \Cref{tab:main} average a Result of 11.2 (median
3.4), and 20 of them score zero. The highest-scoring builder, DeepSeek-V4-Pro,
averages 19.1 yet scores below 20 on eight of ten tasks. Only three runs
exceed 50, all on formal theorem proving or repository bug repair, and no
builder exceeds 30 on more than one task. Success on one commission does not
carry over to the next, and none of the evaluated builders reliably builds
the agents that AgentSWE specifies.

\paragraph{Editing yields partial capabilities, or no delivery at all.}
The builders that deliver on nearly every task, DeepSeek-V4-Pro, GPT-5.5,
and GPT-5.6 Sol, still realize only part of the commissioned capabilities:
their means lie between 20 and 35, and only one run, GPT-5.5 on the Codex
task, reaches 70. Qwen3.6 fails earlier: it submits no candidate on any task,
typically ending its session right after announcing a plan.
DeepSeek-V4-Flash delivers on four tasks; on most of the others it works
until the budget runs out without ever submitting a candidate.
Adding capabilities to a working product thus fails in two ways: the patch
covers only part of the requirement, or the builder never finishes one.

\paragraph{Optimization gains are inconsistent across tasks.}
No builder closes more than 5\,\% of the remaining gap to the target on
average. Of the 25 runs, 13
score above the starter, nine below it, and three match it; under a
per-run sign test, five runs differ from their starter: three BrowseComp
gains and two losses, on $\tau^3$ and PinchBench (\Cref{app:opt-full}). Gains depend on the task: on BrowseComp
four of five builders score above the starter, whereas on PinchBench four
of five score below it.

\subsection{Trajectory Analysis and Failure Modes}
\label{sec:exp-analysis}

Analyzing the builders' trajectories, the frozen artifacts, and the
delivered agents' execution traces in the main runs, we find three
recurring failure modes (\Cref{app:audit} gives the protocol).

\begin{figure}[t]
\centering
\begin{minipage}[t]{0.32\linewidth}
\centering
\includegraphics[width=\linewidth]{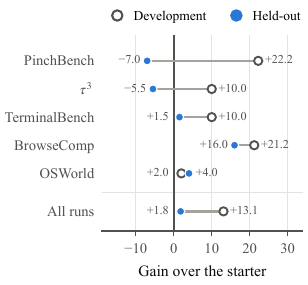}
\caption{Optimization: gain over the starter on development (open) and
held-out (filled) cases, averaged over five builders per benchmark.}
\label{fig:dev-vs-heldout}
\end{minipage}\hfill
\begin{minipage}[t]{0.32\linewidth}
\centering
\includegraphics[width=\linewidth]{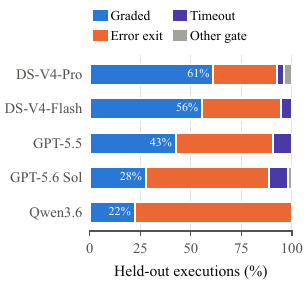}
\caption{Outcomes of held-out Creation executions by builder. Only graded
executions reach the Result rubric.}
\label{fig:creation-outcomes}
\end{minipage}\hfill
\begin{minipage}[t]{0.32\linewidth}
\centering
\includegraphics[width=\linewidth]{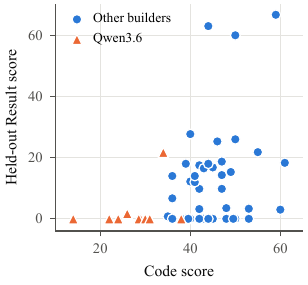}
\caption{Code score against held-out Result score for the 50 Creation
runs.}
\label{fig:code-vs-result}
\end{minipage}
\end{figure}

\paragraph{Overfitting to the visible cases.}
Development scores overstate what a delivery does on the rest of the
requirement, so the gap $F_D - F_{\bar D}$ of \Cref{sec:formulation-problem}
is wide. Eleven Creation runs end development with a score of at least
90, yet their held-out Result averages 9.9; one of them hard-codes its two
development cases (\Cref{app:case}). A DeepSeek-V4-Pro Editing delivery
scores 100 on development cases and 28.3 on held-out cases. In
Optimization, the selected candidate gains 13.1 points on development cases
but only 1.8 on held-out cases (\Cref{app:opt-development}), although both
splits come from one instance pool; on four of the five benchmarks the
gains shrink or reverse (\Cref{fig:dev-vs-heldout}).

Trajectories and artifacts reveal case-specific implementations:
builders fit their agents to the visible cases in 14 of 25 Optimization
runs and 13 of the 33 Editing runs that deliver. On PinchBench, DeepSeek-V4-Flash and
DeepSeek-V4-Pro route tasks by keyword to playbooks for visible task
categories, gaining 8.1 and 59.4 points on development cases and losing
15.2 and 15.1 on held-out cases (\Cref{app:case}). Even explicit feedback leaves such defects
unresolved: in eight of ten DeepSeek-V4-Pro Editing runs, development
feedback names a violated requirement that the delivery still violates on
held-out cases (\Cref{app:edit-audit}). The requirement states what the
held-out cases test; what fails is translating a requirement with
interacting capabilities into an agent that generalizes beyond the visible
cases.

\paragraph{Hollow runtimes.}
Code that reads as complete often does not work when run. In Creation the
failure is loud: 157 of the 270 traced held-out executions (58.1\,\%) fail
a validity gate, nearly all by exiting with an error or timing out
(\Cref{fig:creation-outcomes}). Yet a static review of every frozen
Creation source, by the same judge as Result on an eight-dimension rubric
(\Cref{app:code}), does not separate deliveries with zero Result from the
rest (\Cref{fig:code-vs-result}): apart from Qwen3.6, lowest on both,
deliveries with zero Result score 44.2 and those with nonzero Result 45.8.

In Editing, 28 of the 33 runs that deliver leave some added capabilities
absent, uninvoked, or incompletely executed (\Cref{app:edit-audit,app:case}). These
failures can be silent: an OpenWiki delivery passes every validity gate yet
leaves four of its five required output categories empty on every held-out
case.

\paragraph{Unsupported claims of completion.}
Reports of success are unreliable at both levels of the delegation.
Checked against the builder's own trajectory, artifact, and feedback, 6 of
the 42 final Creation reports and 3 of the 19 Optimization reports claim
completions that these records contradict (\Cref{tab:completion-audit};
\Cref{app:case}).
All nine assert completion; no report that admits partial failure is
contradicted. Delivered Editing agents overclaim at run time too: in 13 of the 33
runs that deliver, they report outcomes that their own effect logs or state
contradict, and the OpenWiki delivery above declares its run complete.

\subsection{Discussion}
\label{sec:exp-discussion}

The failure modes of \Cref{sec:exp-analysis} come from single builds under
one set of conditions. We change these conditions one at a time, namely the
number of development cases, a prompt prior on how to build, the case
budget, and the harness, and we also repeat the build under fixed
conditions, to see how held-out Result and the failure modes respond
(\Cref{app:discussion-protocols}).

\begin{table}[t]
\centering
\scriptsize
\begin{minipage}[t]{0.43\linewidth}
\centering
\caption{Development cases: mean (sd) Result over three builder runs per
condition on the four held-out cases shared by all conditions. Builder:
Codex with DeepSeek-V4.1-Flash.}
\label{tab:devcount}
\setlength{\tabcolsep}{3pt}
\begin{adjustbox}{max width=\linewidth}
\begin{tabular}{@{}lrrrr@{}}
\toprule
\textbf{Task} & \textbf{0 cases} & \textbf{1 case} & \textbf{2 cases} & \textbf{4 cases} \\
\midrule
PDF & 0.0 {\scriptsize(0.0)} & 21.7 {\scriptsize(18.2)} & 34.3 {\scriptsize(6.0)} & 33.2 {\scriptsize(28.8)} \\
PPTX & 4.2 {\scriptsize(7.2)} & 15.0 {\scriptsize(2.8)} & 33.5 {\scriptsize(15.0)} & 15.2 {\scriptsize(1.8)} \\
\bottomrule
\end{tabular}
\end{adjustbox}

\vspace{6pt}
\caption{Prompt priors: mean (sd) Result over three builder runs on six
held-out cases, with the builder of \Cref{tab:devcount}.}
\label{tab:prior}
\setlength{\tabcolsep}{3pt}
\begin{adjustbox}{max width=\linewidth}
\begin{tabular}{@{}lrrr@{}}
\toprule
\textbf{Task} & \textbf{No prior} & \textbf{Strategy A} & \textbf{Strategy B} \\
\midrule
PDF & 39.2 {\scriptsize(14.2)} & 40.4 {\scriptsize(9.3)} & 48.9 {\scriptsize(12.6)} \\
PPTX & 29.6 {\scriptsize(12.6)} & 7.4 {\scriptsize(6.9)} & 31.2 {\scriptsize(4.5)} \\
\bottomrule
\end{tabular}
\end{adjustbox}
\end{minipage}\hfill
\begin{minipage}[t]{0.54\linewidth}
\centering
\caption{Pairings of harness and model, one run each. Creation columns are
held-out Result; Optimization columns are $S$. Codex rows repeat
\Cref{tab:main}; dashes mark pairings outside the study. Sources: DSH (DeepSeek
Harness)~\citep{deepseek-harness2026}, OpenCode~\citep{anomaly2026opencode},
Pi~\citep{earendilworks2026pi}, and GLM-5.3~\citep{glm5team2026glm5vibecodingagentic}.}
\label{tab:harness}
\setlength{\tabcolsep}{3pt}
\begin{adjustbox}{max width=\linewidth}
\begin{tabular}{@{}llrrrr@{}}
\toprule
& & \multicolumn{2}{c}{\textit{Creation}} & \multicolumn{2}{c}{\textit{Optimization}} \\
\cmidrule(lr){3-4}\cmidrule(l){5-6}
\textbf{Harness} & \textbf{Model} & \textbf{GUI} & \textbf{PDF} & $\boldsymbol{\tau^3}$ & \textbf{PinchB.} \\
\midrule
Codex & GPT-5.6 Sol & 18.0 & 18.3 & 4.3 & $-$14.9 \\
Codex & GPT-5.5 & 3.2 & 21.8 & 0.0 & $-$11.1 \\
Codex & DS-V4-Pro & 17.5 & 25.3 & 15.8 & $-$38.7 \\
Codex & DS-V4-Flash & 12.0 & 26.0 & $-$50.0 & $-$34.4 \\
\midrule
DSH & DS-V4-Pro & 19.7 & 20.3 & $-$28.6 & $-$39.7 \\
DSH & DS-V4-Flash & 18.7 & 30.5 & 4.8 & $-$1.7 \\
Claude Code & DS-V4-Pro & 16.2 & 15.3 & -- & -- \\
Claude Code & DS-V4-Flash & 12.7 & 5.8 & -- & -- \\
Claude Code & GLM-5.3 & 0.0 & 19.2 & $-$19.0 & 13.0 \\
OpenCode & GPT-5.6 Sol & 0.0 & 26.7 & -- & -- \\
OpenCode & GPT-5.5 & 0.0 & 42.8 & -- & -- \\
Pi & GPT-5.6 Sol & 0.0 & 18.3 & $-$4.8 & 33.8 \\
Pi & GPT-5.5 & 0.0 & 0.0 & -- & -- \\
\bottomrule
\end{tabular}
\end{adjustbox}
\end{minipage}
\end{table}

\paragraph{Scores rise from zero to two development cases, not beyond.}
Development cases are what builders overfit to (\Cref{sec:exp-analysis}).
On two Creation tasks, with three builder runs per condition, we give the
builder zero, one, two, or four of them. Going from
zero to two development cases raises mean Result by 34.3 points on PDF
translation and by 29.3 on PPTX generation (\Cref{tab:devcount}); without
any case, all three PDF deliveries score zero. Two further cases add nothing
on PDF and lower PPTX from 33.5 to 15.2 (\Cref{fig:repeat-runs}).

\paragraph{A prescribed design lowers PPTX; self-testing narrows its spread.}
We also append one of two prompt priors to the builder instruction
(\Cref{app:priors}). Strategy A prescribes a design, the components and
how they connect; Strategy B asks the builder to write additional cases and
run its agent end to end, aiming at the hollow runtimes and unsupported
claims of \Cref{sec:exp-analysis}. Strategy A leaves PDF unchanged
and lowers PPTX from 29.6 to 7.4; Strategy B raises PDF from 39.2 to 48.9,
within run-to-run variation, and leaves the PPTX mean unchanged while
narrowing its spread (\Cref{tab:prior}).

\paragraph{The gap remains with 2.5 times the case budget.}
Only 15 of the 270 held-out Creation executions in the main runs reach the
600\,s limit (\Cref{fig:creation-runtime}). Rebuilding all ten Creation
tasks with 1{,}500\,s per case and a requirement for a traced runtime agent
(\Cref{app:discussion-protocols}) leaves both failure modes in place: no
execution reaches the limit, the last development score exceeds held-out
Result on every task, 70.3 against 31.5 on average (\Cref{tab:headroom}),
and a third of the held-out executions produce no valid result.

\paragraph{Scores vary across harness--model pairings.}
With the model held fixed, scores differ across harnesses
(\Cref{tab:harness}): on scientific PDF translation, GPT-5.5 scores 21.8
under Codex and 42.8 under OpenCode, and DeepSeek-V4-Flash 5.8 under
Claude Code and 30.5 under DSH. No harness leads for every model, so we
evaluate pairings of harness and model, not models alone.

\begin{table}[t]
\centering
\caption{AgentSWE-Lite: mean over three builds per pairing in Codex,
held-out Result for Creation and Editing and $S$ for Optimization;
\Cref{fig:lite-repeats} shows each build. DS-V4.1-Flash (DeepSeek-V4.1-Flash)
is the runtime model and the judge in all three families
(\Cref{app:discussion-protocols}), so Creation and Optimization levels are
not comparable to \Cref{tab:main}.}
\label{tab:lite}
\small
\setlength{\tabcolsep}{4pt}
\begin{tabular}{@{}lrrr@{}}
\toprule
\textbf{Task} & \textbf{GPT-5.6 Sol} & \textbf{DS-V4.1-Flash} & \textbf{Qwen3.6} \\
\midrule
\multicolumn{4}{@{}l}{\textit{Creation}} \\
Scientific PDF translation & 18.9 & 31.9 & 0.0 \\
Database analytics & 7.4 & 38.7 & 0.0 \\
Repository bug repair & 42.9 & 30.0 & 0.0 \\
\quad Mean & 23.1 & 33.5 & 0.0 \\
\midrule
\multicolumn{4}{@{}l}{\textit{Editing}} \\
DeepTutor adaptive remediation & 52.2 & 27.6 & 0.0 \\
Aider worktree transaction & 13.0 & 39.3 & 6.0 \\
OpenWiki change-impact documentation & 22.8 & 21.0 & 0.0 \\
\quad Mean & 29.3 & 29.3 & 2.0 \\
\midrule
\multicolumn{4}{@{}l}{\textit{Optimization}} \\
$\tau^3$ & 10.1 & 13.0 & 4.3 \\
PinchBench & 19.4 & 24.3 & 2.4 \\
\quad Mean & 14.8 & 18.7 & 3.4 \\
\bottomrule
\end{tabular}
\end{table}

\paragraph{The same pairing delivers very different agents across builds.}
The main comparison and harness studies build each pairing once. On eight
tasks from the three families, we build each of three builders three times
under fixed conditions (\Cref{tab:lite}; \Cref{app:discussion-protocols}).
Builds of one pairing can lie far apart (\Cref{fig:lite-repeats}): GPT-5.6
Sol scores 32.8, 33.3, and 90.5 on DeepTutor, and DeepSeek-V4.1-Flash 0.0,
33.3, and 56.8 on repository bug repair. No builder leads on every task, and
Qwen3.6 averages at most 6.0 on any task. Overfitting recurs: two GPT-5.6
Sol builds of database analytics reach 99 on the development cases and 11.0
and 6.5 on held-out cases (\Cref{sec:exp-analysis}). The repository provides
these eight tasks and this configuration as AgentSWE-Lite, a lower-cost entry point for evaluation.

\section{Conclusion}
\label{sec:conclusion}

We introduced AgentSWE to measure whether coding agents deliver agents that
serve their requirement on cases they have not seen. Current builders
rarely do: in the main comparison, the strongest average 19.1 on Creation
and 31.6 on Editing, and
in Optimization none closes more than 5\,\% of the remaining gap to the
target on average. The trajectories reveal recurring failures: builders fit the visible
cases, ship runtimes that read as complete but do not work, and, like the
agents they deliver, claim completions their own records contradict. These
failure modes chart where coding agents need to improve; our future work will
pursue harnesses that test against the requirement and verify completion
claims, and training that rewards behavior on unseen cases.

\subsection*{AI Use Statement}
We used generative AI tools for three tasks. First, to construct the
benchmark, we used Codex with GPT-5.6 Sol to synthesize task materials from
the analyzed source systems (\Cref{sec:benchmark-sourcing}): the
requirement documents, the development and held-out cases, and the
task-specific Result rubrics. The authors designed each commission and
reviewed and revised the generated materials. Second, AI agents assisted
the trajectory analysis (\Cref{sec:exp-analysis,app:audit}) by labeling
failure modes from builder trajectories, execution traces, and evaluator
records; the authors reviewed the labels and the resulting counts. Third,
we used AI assistants to aid and polish the writing of this paper. Coding agents and
language models are also the object of study, acting as builders, runtime
models, and judges (\Cref{sec:exp-setup,app:setup}); this use is part of
the method rather than assistance to the authors. We have reviewed all
AI-assisted content and take full responsibility for the paper.

\subsection*{Ethics Statement}
AgentSWE asks coding agents to build agents that execute code, access the
web, and, in one Creation task, validate software vulnerabilities. All
builder sessions and candidate executions run in isolated containers with
pinned fixtures, and every model call passes through a benchmark-owned
gateway (\Cref{app:infra}). The benchmark involves no human subjects. Its
tasks derive from public repositories and public benchmarks, used under
their licenses and terms. Our analysis documents how coding agents
misreport their own work, which we regard as a safety-relevant property of
delegating software to them.

\subsection*{Reproducibility Statement}
\Cref{sec:benchmark} and \Cref{app:tasks,app:rubrics} describe the tasks,
cases, and scoring. \Cref{app:infra,app:setup} give the evaluation
infrastructure, builder instructions, model versions, effort settings, and
budgets. \Cref{app:noise} reports repeated runs and the variation of
repeated evaluation, \Cref{app:results} lists per-run results, and
\Cref{app:audit,app:case} describe the analysis protocol and the traced
cases. The benchmark, comprising the task packages, the development and
held-out cases, the Creation and Editing rubrics and judge prompts, the
evaluators, and the evaluation harness, is available at \coderepo. Where an
upstream license does not permit redistribution, the repository fetches the
pinned upstream commit instead of including it. The repository also
provides the eight tasks and the configuration of \Cref{tab:lite} as
AgentSWE-Lite.

\bibliographystyle{iclr2027_conference}
\bibliography{citation}

\clearpage
\appendix
\section{Task Sources and Statistics}
\label{app:task-sources}

\Cref{tab:task-sources} lists the 25 tasks with the system each one derives
from and its numbers of development and held-out cases.  Creation tasks
reverse-engineer the listed domain agent into a requirement, Editing tasks
extend the pinned codebase of the listed product, and Optimization tasks
keep the native evaluator of the listed benchmark and start from the listed
agent.

\begin{table}[h]
\centering
\caption{AgentSWE tasks, their source systems, and their development and
held-out cases.}
\label{tab:task-sources}
\footnotesize
\setlength{\tabcolsep}{4pt}
\begin{tabularx}{\linewidth}{@{}l>{\raggedright\arraybackslash}X>{\raggedright\arraybackslash}Xr@{}}
\toprule
\textbf{Family} & \textbf{Task} & \textbf{Source system} & \textbf{Dev / held-out} \\
\midrule
Creation & Authorized vulnerability validation & Strix~\citep{usestrix2026strix} & 2 / 6 \\
 & Database analytics & Vanna~\citep{vannaai2026vanna} & 2 / 6 \\
 & Desktop GUI automation & UI-TARS-desktop~\citep{qin2025uitars} & 2 / 6 \\
 & Document-to-editable PPTX & PPT Master~\citep{he2026pptmaster} & 2 / 6 \\
 & Evidence-grounded document QA & PaperQA~\citep{lala2023paperqa,skarlinski2024paperqa2} & 2 / 6 \\
 & Formal theorem proving & LeanCopilot~\citep{song2025leancopilot} & 2 / 6 \\
 & Repository bug repair & SWE-agent~\citep{yang2024sweagent} & 2 / 6 \\
 & Schema-guided web extraction & ScrapeGraphAI~\citep{Perini_Scrapegraph-ai_2024} & 2 / 6 \\
 & Scientific PDF translation & PDFMathTranslate~\citep{ouyang2025pdfmathtranslate} & 2 / 6 \\
 & Web research report & MiroFlow~\citep{su2026miroflow} & 2 / 6 \\
\midrule
Editing & Claude policy provenance & Claude Code plugins~\citep{anthropic2026claudecode} & 2 / 6 \\
 & Aider worktree transaction & Aider~\citep{gauthier2026aider} & 2 / 6 \\
 & OpenHands effect recovery & OpenHands~\citep{wang2024openhands} & 2 / 6 \\
 & OpenClaw channel handoff & OpenClaw~\citep{openclawcontributors2026openclaw} & 2 / 6 \\
 & Codex execution residual & Codex CLI~\citep{openai2026codexcli} & 2 / 6 \\
 & AI-Scientist reproducibility gate & AI-Scientist-v2~\citep{yamada2025aiscientistv2} & 2 / 6 \\
 & DeepCode claim traceability & DeepCode~\citep{li2025deepcode} & 2 / 6 \\
 & DeepTutor adaptive remediation & DeepTutor~\citep{zhao2026deeptutor} & 2 / 6 \\
 & Dyad acceptance-driven workflow & Dyad~\citep{dyadsh2026dyad} & 2 / 6 \\
 & OpenWiki change-impact documentation & OpenWiki~\citep{sproul2026openwiki} & 2 / 6 \\
\midrule
Optimization & BrowseComp~\citep{wei2025browsecomp} & Search ReAct agent~\citep{yao2023react} & 50 / 100 \\
 & $\tau^3$~\citep{yao2024taubench,barres2025tau2bench} & Policy-aware ReAct agent~\citep{yao2023react} & 20 / 40 \\
 & TerminalBench~\citep{merrill2026terminalbench} & Terminus-2 agent~\citep{merrill2026terminalbench} & 10 / 40 \\
 & PinchBench~\citep{kiloai2026pinchbench} & OpenClaw productivity agent & 20 / 40 \\
 & OSWorld~\citep{xie2024osworld} & Screenshot ReAct policy~\citep{yao2023react} & 10 / 25 \\
\midrule
Total & 25 tasks & & 150 / 365 \\
\bottomrule
\end{tabularx}
\end{table}

\section{Representative Tasks and Case Coverage}
\label{app:tasks}
\label{app:example-task}

Each card below shows, for one representative task per family, what the
builder receives: the section structure of the four requirement documents
with excerpts, one public development case, and a summary of the scoring.
Excerpts are abridged, with omissions marked [\ldots]; the builder never sees
the held-out cases. \Cref{tab:case-coverage} contrasts the requirements that
development and held-out cases exercise.

\definecolor{cardReq}{HTML}{2A78D6}
\definecolor{cardDev}{HTML}{1BAF7A}
\definecolor{cardScore}{HTML}{4A3AA7}
\tcbset{taskcard/.style={enhanced, breakable, boxrule=0.6pt, arc=2pt,
  left=6pt, right=6pt, top=4pt, bottom=4pt, fontupper=\small,
  coltitle=white, fonttitle=\small\bfseries, toptitle=1pt, bottomtitle=1pt,
  before skip=6pt, after skip=6pt},
  reqcard/.style={taskcard, colframe=cardReq, colbacktitle=cardReq, colback=cardReq!4!white},
  devcard/.style={taskcard, colframe=cardDev, colbacktitle=cardDev, colback=cardDev!5!white},
  scorecard/.style={taskcard, colframe=cardScore, colbacktitle=cardScore, colback=cardScore!4!white}}
\newcommand{\taskdoc}[2]{\par\noindent\textbf{\texttt{#1}}\enspace{\footnotesize\color{black!65}#2}\par\smallskip}

\subsection{Creation: Scientific PDF Translation}

The builder starts from an empty workspace and must deliver a reusable
translator with an offline viewer that reads in both directions.

\begin{tcolorbox}[reqcard, title={Requirement \,\textbar\, four Markdown documents}]
\taskdoc{01\_task\_goal.md}{Background $\cdot$ Agent to Create $\cdot$ Target Users $\cdot$ Core Value $\cdot$ Final Objective $\cdot$ Non-Goals}
``You must create an \textbf{Interactive Scientific PDF Translation Agent}. It
receives one Markdown request that identifies one or more local scientific
PDFs, source and target languages, a translated-PDF display mode, and
optional terminology assets. It produces a faithful translated PDF plus a
self-contained local reading viewer that displays the source and translation
side by side with geometry-based paragraph correspondence.''
\medskip

\taskdoc{02\_interface\_and\_delivery.md}{Environment Preparation $\cdot$ Uniform Run Command $\cdot$ Input Contract $\cdot$ Output Directory and Repeated Runs $\cdot$ Required Successful Output $\cdot$ PDF Contract $\cdot$ Alignment Artifact $\cdot$ Viewer Interaction Contract $\cdot$ \texttt{run\_report.json} $\cdot$ Exit Codes and Failure Reports}
``Every case is invoked exactly as \texttt{python run\_agent.py --input
<input.md> --output <output\_dir>} [\ldots] A successful run must contain
all of these paths:'' \texttt{translated.pdf}, \texttt{alignment.json},
\texttt{run\_report.json}, \texttt{viewer/index.html}, \texttt{styles.css},
\texttt{app.js}, \texttt{documents/source.pdf},
\texttt{documents/translated.pdf}, and \texttt{vendor/}. ``The browser must
make no external network request after the output directory has been
generated.''
\medskip

\taskdoc{03\_requirements\_and\_constraints.md}{nine functional requirements: interpret the complete request $\cdot$ inspect every page and recover reading order $\cdot$ translate faithfully $\cdot$ preserve scientific objects and associations $\cdot$ produce a readable translated PDF $\cdot$ build geometry-grounded alignment $\cdot$ build the local viewer $\cdot$ validate before success $\cdot$ degrade honestly; implementation constraints; observable evaluation boundary}

\taskdoc{04\_resources.md}{Evaluator-owned model transport $\cdot$ Environment Ownership $\cdot$ Runtime Envelope $\cdot$ Credentials $\cdot$ Multimodal Responses API $\cdot$ API and Retrieval Budgets $\cdot$ Serper-Compatible Search $\cdot$ Search/Scrape Tool Layer $\cdot$ Network and Filesystem}
``Wall-clock limit: 600 seconds per case, including translation and
validation. Memory limit: 4 GiB per case. Generation is non-interactive.''
\end{tcolorbox}

\begin{tcolorbox}[devcard, title={Development case \texttt{dev\_002} \,\textbar\, Thermal Barrier Imaging Note}]
Translate \texttt{source\_mixed.pdf} from English to Simplified Chinese for
materials-science researchers. Apply \texttt{glossary.csv} as authoritative
and create the complete interactive reading bundle.

\textit{Source condition.} The four-page PDF is mixed: pages 1, 2, and 4
contain native selectable text, while page 3 is an embedded scan with mild
skew, faint type, a plot, and the printed batch mark \texttt{TB-17}. Page 4
is landscape. Recover page 3 by OCR or visual analysis without inventing
illegible characters.

\textit{Translated PDF.} Produce a \textbf{bilingual English-Simplified
Chinese PDF} with exactly four pages and the same
portrait/portrait/portrait/landscape sequence. Within each prose region,
place the English source first and its Chinese translation directly below.
[\ldots] Follow every glossary target exactly. Preserve \texttt{TB-17},
coating code \texttt{YSZ-8}, instrument \texttt{IRX-4}, citation markers,
signs, decimal precision, units, $\alpha = k/(\rho c_p)$,
$\mathrm{Fo} = \alpha t/L^2$, and equation numbers (1) and (2).

\textit{Alignment and viewer.} Align source paragraphs to the Chinese part
of each bilingual target region, not to the duplicated English. Include
OCR-derived page-3 paragraphs with honest confidence. Visibly distinguish
low-confidence links in the viewer. Bidirectional activation must work on
native page 2, scanned page 3, and rotated landscape page 4. The supplied
assets are the complete authority. Make zero search or retrieval calls and
add no scan disclaimer page.
\end{tcolorbox}

\begin{tcolorbox}[scorecard, title={Scoring}]
Result rubric with six dimensions: semantic completeness and terminology 35,
fidelity of scientific objects and data 35, page and reading structure 12,
alignment auditability 8, offline viewer 8, delivery integrity and
accessibility 2; gates and ceilings as described in \Cref{app:rubrics}.
The frozen source is also graded once on the shared static rubric
(\Cref{app:code}).
\end{tcolorbox}

The difficulty is compositional: a correct translation still fails part of
the commission if its alignment points to the duplicated English, or if the
viewer cannot find the rendered region.

\subsection{Editing: OpenHands Effect Recovery}

The builder receives a pinned OpenHands product and must add durable
recovery through its production services, delivering a patch with edit and
run reports.

\begin{tcolorbox}[reqcard, title={Requirement \,\textbar\, four Markdown documents and a checkpoint schema}]
\taskdoc{01\_task\_goal.md}{Task goal: cross-instance transactional effect delivery}
``Extend the supplied OpenHands Agent Canvas repository with a
backend-scoped, versioned recovery ledger for conversation effects. The
product must stay honest when two service instances or browser tabs share
storage, a process stops after any delivery stage, a lease expires and moves
to a new runtime, or pause/resume/cancel races an old callback. The ledger is
a transactional outbox, not merely a history snapshot.''
\medskip

\taskdoc{02\_interface\_and\_delivery.md}{Submission $\cdot$ Frozen TypeScript adapter $\cdot$ Recovery-aware conversation event dispatcher $\cdot$ Durable cross-tab recovery sync coordinator $\cdot$ Transactional workspace recovery reconciler $\cdot$ Persistence, integrity, and migration $\cdot$ Delivery and crash semantics $\cdot$ Lease, ABA, event, and compaction semantics $\cdot$ Least-privilege inspection and production seams $\cdot$ Error contract for deterministic crash injection}
``Submit exactly these top-level files: \texttt{solution.patch}: a non-empty
UTF-8 unified Git patch relative to the supplied repository that applies
with \texttt{git apply} exactly once. \texttt{edit\_report.json} [\ldots]
\texttt{run\_report.json} [\ldots] Reports are evidence, not assertions of
correctness.''
\medskip

\taskdoc{03\_requirements\_and\_constraints.md}{functional requirements $\cdot$ constraints $\cdot$ verifiable hard requirements}
``Implement the durable outbox state machine: \texttt{queued ->{} claimed ->{}
applied ->{} completed}, with \texttt{uncertain}, \texttt{blocked}, and
\texttt{cancelled} recovery outcomes. Persist every stage before continuing,
and never infer callback success from a callback return digest alone.''
[\ldots] ``Modify only \texttt{src/api/conversation-service/**},
\texttt{src/api/runtime-service/**}, \texttt{src/api/recovery/**},
\texttt{src/stores/**} [\ldots] Do not modify lockfiles, package/build
configuration, benchmark cases, generated output, authentication, network
policy, or unrelated product code.''
\medskip

\taskdoc{04\_resources.md}{Pinned runtime $\cdot$ Prewarm and install $\cdot$ Runtime and memory envelope $\cdot$ Filesystem}
\end{tcolorbox}

\begin{tcolorbox}[devcard, title={Development case \texttt{dev\_001} \,\textbar\, Converge an interrupted workspace and a lossy history stream}]
Resume this public practice conversation. A chunk write response and a
transport pull response were both lost after the far side had already acted
on them, the history channel echoes its own notices back, and the workspace
has a local-only and a remote-only path to reconcile. Five outcomes are
required, and they are scored independently.
\begin{enumerate}[nosep,leftmargin=*,label=\textbf{\arabic*.}]
\item \textbf{Upload exactly the missing chunk, once}, after re-probing what
  the remote side genuinely lacks.
\item \textbf{Converge exactly once} on the union of the local-only and
  remote-only paths: one durable remote commit, one local apply.
\item \textbf{Retry the lost transport response from the same durable
  cursor}, not from the advisory remote cursor.
\item \textbf{Complete, bounded projection; no duplicate effect.} No side
  effect runs twice, and the public workspace projection carries every
  published stage field and no path, digest, chunk identifier, tab or
  incarnation identity, or token. An empty projection is not a safe
  projection.
\item \textbf{Keep runtime-event discipline.} A delivery that carries the
  current owner, lease epoch, fencing token, run generation and expected
  revision is genuine: accept it [\ldots] A delivery whose fencing token or
  lease epoch does not match the live authority must be suppressed [\ldots]
  Suppressing everything is not a safe answer.
\end{enumerate}
Report what actually happened. Claiming an outcome the product did not reach
scores worse than reporting it honestly as unmet.
\end{tcolorbox}

\begin{tcolorbox}[scorecard, title={Scoring}]
A patch that does not apply scores zero. Otherwise the Result rubric
scores task completion 35, recovery quality
25, durable consistency 20, honest reporting 10, and privacy and scope 10,
from numbered outcomes with observed facts, satisfaction flags, and safety
flags; unmet outcomes impose ceilings (\Cref{app:rubrics} reconstructs one
recorded score).
\end{tcolorbox}

The commission pairs a positive obligation to resume legitimate work with a
negative obligation to reject stale work; an agent that refuses every action
fails both.

\subsection{Optimization: PinchBench Productivity Agent}

The builder receives a runnable OpenClaw productivity agent and improves it
against the native PinchBench objective. The requirement is short enough to
show almost in full.

\begin{tcolorbox}[reqcard, title={Requirement \,\textbar\, four Markdown documents and a task contract}]
\taskdoc{01\_task\_goal.md}{Goal $\cdot$ Non-goals}
``You are optimizing the runnable OpenClaw Basic Productivity Agent already
installed in the submission workspace. [\ldots] Build a reusable tool-using
productivity agent that completes file, calendar, analysis, browser,
research, and codebase tasks in a controlled workspace, including recovery
and cross-step planning.'' Non-goals: hard-coding the development cases or
hidden answers, self-reported rewards, and copying a complete third-party
agent or evaluator.
\medskip

\taskdoc{02\_interface\_and\_delivery.md}{one executable harness that emits an OpenClaw agent specification per input row}
{\footnotesize
\begin{verbatim}
python run_harness.py --input <input.jsonl> \
    --output <predictions.jsonl> --run-dir <run_dir>

{"id": "dev_001",
 "agent": {"kind": "openclaw",
           "instructions": "reusable system-level working policy",
           "planning_guidance": "optional task planning policy",
           "recovery_guidance": "optional failure recovery policy",
           "verification_guidance":
             "optional artifact verification policy",
           "tool_profile": "coding"}}
\end{verbatim}
}
``The evaluator locks the OpenClaw/model/provider versions, task prompt,
workspace fixtures, timeout, model-call/token budget, transcript collection,
official grader and hybrid judge. It ignores self-reported success or
rewards.''
\medskip

\taskdoc{03\_requirements\_and\_constraints.md}{functional requirements $\cdot$ implementation constraints}
``The supplied \texttt{run\_harness.py} and \texttt{productivity\_policy.py}
are the only editable paths.'' [\ldots] ``Never access hidden cases,
evaluator gold, reference trajectories, or other submissions.''
\medskip

\taskdoc{04\_resources.md}{pinned runtime and budgets}
``The evaluator runs the submitted specification with pinned OpenClaw
\texttt{2026.7.1-2} [\ldots] Budget per row: at most 20 model calls, 600
seconds, and 200,000 model tokens. [\ldots] Automated and hybrid scoring use
the pinned PinchBench \texttt{grade\_task()} implementation.''
\end{tcolorbox}

\begin{tcolorbox}[devcard, title={Development case \texttt{dev\_007} \,\textbar\, \texttt{task\_log\_apache\_top\_errors}}]
Optimization Task \texttt{task\_log\_apache\_top\_errors}. This task is
executed by the pinned evaluator-owned OpenClaw runtime. Optimize a reusable
productivity-agent policy; the official task prompt, fixtures, workspace,
transcript and \texttt{grade\_task()} remain evaluator-owned.
\end{tcolorbox}

\begin{tcolorbox}[scorecard, title={Scoring}]
The pinned native grader scores each PinchBench task, automatically or with
the benchmark's hybrid judge. The starter and the selected candidate run on
the same 40 held-out tasks, and the task reports gap closure $S$
(\Cref{eq:opt}); the 20 development tasks provide feedback during the five
accepted rounds.
\end{tcolorbox}

The builder edits a reusable policy; the objective measures the work the
runtime completes under it.

\begin{table}[t]
\centering
\caption{Requirement coverage in development and held-out cases, with
examples. The held-out split changes inputs and interactions within the
commission.}
\label{tab:case-coverage}
\footnotesize
\setlength{\tabcolsep}{4pt}
\begin{tabularx}{\linewidth}{@{}p{2.0cm}XX@{}}
\toprule
\textbf{Requirement} & \textbf{Public development example} & \textbf{Held-out example} \\
\midrule
PDF content and geometry & Mixed native and scanned pages, bilingual Chinese output, a landscape page (dev 002). & English-to-Japanese translation of a two-column paper with a paragraph split across pages 2 and 3 (test 002); Traditional Chinese with dense figure labels and preferred glossary terms (test 006). \\
OpenHands recovery & Lost responses, convergence applied exactly once, discipline for runtime events (dev 001); fencing of the replaced owner and scoped access (dev 002). & Pause, resume, and cancel followed by stale callbacks and a terminal burst (test 003); durable recovery across three workspace crash boundaries (test 002). \\
PinchBench transfer & Stock trend analysis (dev 006) and Apache top error analysis (dev 007). & Stock volatility analysis (test 004), Apache critical errors and timelines (tests 013 and 014), under the same fixed runtime and native grader. \\
\bottomrule
\end{tabularx}
\end{table}

\section{Result Rubrics and a Worked Score}
\label{app:rubrics}

Each Result rubric scores the observable outcomes of its task. We describe
two of them and reconstruct one recorded score.

\paragraph{PDF translation: content dominates presentation.}
The six Result dimensions allocate 35 points to semantic completeness and
terminology, 35 to fidelity of scientific objects and data, 12 to page and
reading structure, 8 to alignment auditability, 8 to the offline viewer,
and 2 to delivery integrity and accessibility, so 70 points concern the
translated content. The evaluator checks the final bundle, the rendered
pages and their geometry, and the viewer in a browser; source architecture
earns no Result credit. A run that does not launch, exceeds a resource
limit, or produces no parseable translated PDF scores zero. A valid but
incomplete translation is still scored, under ceilings: omitting or leaving
untranslated at least 25\,\% of the substantive units caps the total at 35,
a material reversal of meaning caps it at 45, and repeated corruption of
formulas or values caps it at 40. The lowest applicable ceiling bounds the
sum of the dimension scores. A valid PDF without alignment and viewer
artifacts is capped at 75, which separates missing functionality from
execution failure.

\paragraph{OpenHands: outcomes, ceilings, and earned credit.}
The dimensions are task completion (35), recovery quality (25), durable
consistency (20), honest reporting (10), and privacy and scope (10). Each
case lists numbered outcomes with observed facts, satisfaction flags, and
safety flags. Unmet outcomes and safety failures impose ceilings on each
dimension, and all applicable ceilings apply together.

\Cref{tab:worked-score} reconstructs the recorded score of GPT-5.6 Sol's
OpenHands delivery on \texttt{test\_002}. None of the five outcomes is
satisfied: no missing chunk is uploaded, none of the three crash boundaries
is reached, there are no remote commits and no local applies, no restart
demonstrates durability, and the genuine event is not accepted. Four
outcomes are safety requirements. The agent reports that it is blocked and
acknowledges the unmet outcomes, which keeps some honesty credit. The score
is $0+0+1+2+1=4$, although checkpoint and lease setup succeed.

\begin{table}[t]
\centering
\caption{Worked OpenHands Result score. Ceilings for unmet outcomes and
ceilings for safety failures apply together; their minimum per dimension
sums to 10, and the recorded score is 4.}
\label{tab:worked-score}
\small
\setlength{\tabcolsep}{4pt}
\begin{tabularx}{\linewidth}{@{}Xrrrrr@{}}
\toprule
\textbf{Dimension} & \textbf{Max.} & \textbf{Partial cap} & \textbf{Safety cap} & \textbf{Effective cap} & \textbf{Score} \\
\midrule
Task completion & 35 & 4 & 3 & 3 & 0 \\
Recovery quality & 25 & 3 & 2 & 2 & 0 \\
Durable consistency & 20 & 2 & 2 & 2 & 1 \\
Honest reporting & 10 & 2 & 3 & 2 & 2 \\
Privacy and scope & 10 & 1 & 2 & 1 & 1 \\
\bottomrule
\end{tabularx}
\end{table}

\section{Evaluation Infrastructure and Builder Instructions}
\label{app:infra}

Builder development and candidate evaluation run as separate Harbor tasks.
The builder's writable workspace holds the delivery and its permitted
dependencies. The public package exposes the requirement and the
development cases; the hidden cases and the evaluator stay outside it.
Candidate snapshots execute in an environment built for the task, with
pinned fixtures, services, and resource limits, and route model requests
through the evaluator. Optimization also keeps the source benchmark's
native executor and grader. OpenHands runs record both the product's own
observations and a separate outcome ledger, so an audit can compare what
the agent reports with the state it changed.

\paragraph{Base instruction and prompt priors.}
The PDF builder instruction specifies one continuous development loop:
\begin{quote}\small
``Your entire build/dev/repair loop occurs in this one native session and
this one writable workspace; no fresh repair conversation will be created.''

``The returned feedback is the only authoritative dev feedback. Diagnose
it, edit the same \texttt{/workspace/submission}, and submit again in this
same conversation.''
\end{quote}
The prompt priors of \Cref{tab:prior} are appended to the base builder
instruction; their full texts are in \Cref{app:priors}. The feedback the builders
received appears in \Cref{app:case}; runtime settings are in
\Cref{app:setup}.

\subsection{Prompt Priors}
\label{app:priors}

In the prompt-prior conditions of \Cref{tab:prior}, one of the texts below
is appended to the base builder instruction. They are reproduced verbatim;
the private environment path is abbreviated and omissions are marked
\texttt{[...]}.

\definecolor{stratA}{HTML}{2A78D6}
\definecolor{stratB}{HTML}{EB6834}
\tcbset{promptbox/.style={enhanced, breakable, boxrule=0.6pt, arc=2pt,
  left=5pt, right=5pt, top=3pt, bottom=3pt, fontupper=\scriptsize,
  coltitle=white, fonttitle=\small\bfseries, toptitle=1pt, bottomtitle=1pt,
  before skip=8pt, after skip=8pt}}

\begin{tcolorbox}[promptbox, colframe=stratA, colbacktitle=stratA, colback=stratA!4!white,
  title={Strategy A \,\textbar\, Scientific PDF translation}]
\begin{verbatim}
## Strategy A prior: architecture for scientific PDF translation

Treat the created system as a geometry-preserving document translation pipeline, not
as page-level text replacement. Preserve freedom in libraries and models, but make
these architectural responsibilities explicit:

1. Analyze each page before translation. Render and parse it, identify native versus
   scanned content, recover reading order and layout regions, and classify prose,
   formulas, code, tables, figures, captions, footnotes, references, headers, and
   protected identifiers. Use OCR or visual recovery only where native extraction is
   inadequate, with confidence attached to recovered units.
2. Maintain an intermediate document model whose units carry source page, reading
   order, text, geometry, type, protection policy, and confidence. Translation should
   operate on coherent prose units while scientific objects and their associations
   remain explicitly protected.
3. Separate extraction, translation, and reconstruction. Batch or cache translation
   where useful, enforce glossary/protected-token constraints, and validate numbers,
   signs, units, citations, equations, identifiers, negation, and modality before
   placing target text back into the document.
4. Reconstruct adaptively within the original page geometry. Choose suitable fonts,
   script direction, line breaking, local reflow, and bilingual pairing. Preserve
   page count, dimensions, rotations, columns, object associations, and page-local
   reading order while avoiding clipping, overlap, and text painted across graphics.
5. Derive alignment from the same intermediate geometry used for reconstruction.
   Map meaningful source/target units with real page boxes, multiple anchors when a
   unit splits, stable reading-order IDs, digests, status, and calibrated confidence.
   Do not invent decorative boxes or align against a detached transcript.
6. Build the viewer as an offline projection of the PDF and alignment artifacts.
   Bundle the renderer locally; use the alignment geometry for overlays; support
   bidirectional pointer and keyboard activation, active/ARIA synchronization,
   counterpart scrolling, page identity, and visible uncertainty.
7. Validate in layers before success: parse and render every PDF page; compare page
   count/dimensions/rotation; check glyph coverage and protected-token survival;
   validate alignment schema, bounds, order, digests, and visible overlap; then run a
   clean-browser offline interaction smoke at both required viewports.
8. Recover honestly. Retry bounded OCR/translation/layout operations, isolate
   page-local failures, preserve uncertain visible tokens, lower confidence when
   warranted, and never silently omit a page or fabricate an alignment.

The reusable pattern is page analysis -> typed geometry model -> constrained
translation -> adaptive reconstruction -> geometry-derived alignment -> offline
viewer -> parse/render/browser validation and targeted repair.
\end{verbatim}
\end{tcolorbox}

\begin{tcolorbox}[promptbox, colframe=stratB, colbacktitle=stratB, colback=stratB!4!white,
  title={Strategy B \,\textbar\, Scientific PDF translation}]
\begin{verbatim}
## Strategy B prior: self-created development and self-verification

Do not optimize only against `dev_001` and `dev_002`. During this same Builder
session, create a private synthetic suite under `/workspace/self_cases`, run the
current created agent on it, and judge the complete translation bundles yourself.
Keep synthetic cases, outputs, renders, and review notes outside
`/workspace/submission`.

The Builder container provides the same evaluator-owned lower-agent transport used by
official Candidate jobs. [...] Invoke the current agent with the documented
private environment, for example:

    <private-env>/bin/python \
      /workspace/submission/run_agent.py \
      --input /workspace/self_cases/<case>/input.md \
      --output /workspace/self_results/<case>

Create structurally different PDFs rather than paraphrases of the public examples.
Across the session cover native and scan-only pages, multi-column and landscape
layouts, mixed rotations, monolingual and bilingual output, glossary conflicts,
protected equations/identifiers/units, tables/figures/captions/footnotes/references,
long text expansion, low-confidence OCR, split units, and degraded but recoverable
input. Generate small synthetic PDFs and glossaries locally so expected content and
geometry are known.

Before each run, write expected observable invariants. Then parse and render every
output page; compare count/dimensions/rotation and protected tokens; inspect target
glyphs, clipping, overlap, reading order, scientific associations, and glossary
consistency; validate alignment schema/digests/bounds/visible box overlap/confidence;
serve the viewer offline and exercise bidirectional click and keyboard activation at
both viewports. Record results and mechanism-level causes in
`/workspace/self_test_log.md`, then repair general extraction, translation,
reconstruction, alignment, viewer, or validation mechanisms.

Use official submissions as periodic external calibration inside the ten-round
budget: study feedback, update the synthetic coverage matrix, run relevant
self-created cases with the lower agent, self-evaluate, repair, and retain a compact
regression suite. Aim for at least six distinct synthetic cases spanning every major
function. High scores on the two public documents do not prove robustness to new
page structures or languages.
\end{verbatim}
\end{tcolorbox}

\begin{tcolorbox}[promptbox, colframe=stratA, colbacktitle=stratA, colback=stratA!4!white,
  title={Strategy A \,\textbar\, Document-to-editable PPTX}]
\begin{verbatim}
## Strategy A prior: architecture for document-to-editable-PPTX generation

Treat the created system as a staged document-understanding and presentation
compiler, not as a case-specific slide template. Preserve freedom in implementation,
but make these architectural responsibilities explicit:

1. Normalize every authorized source into a provenance-bearing content model. Parse
   text, headings, lists, tables, figures, captions, links, page/section locations,
   and source authority before slide authoring. Keep claims and quantitative values
   tied to source records rather than passing around untraceable prose.
2. Separate content strategy from visual execution. First infer audience, purpose,
   required slide count, narrative arc, slide roles, key claims, and evidence. Then
   create a slide-by-slide plan with one communication objective per slide. Only then
   materialize editable PowerPoint objects.
3. Establish a deck-wide design system before individual slides: canvas, margins,
   grid, typography hierarchy, palette, recurring components, image treatment, and
   density rules. Allow deliberate variation by slide role while preserving visual
   continuity; do not generate disconnected pages independently.
4. Use semantic slide objects. Titles, body text, shapes, tables, charts, diagrams,
   images, captions, citations, and speaker notes should remain natively editable and
   meaningfully grouped. Use native charts/tables when data semantics matter. Do not
   use full-slide screenshots or rasterized text as a shortcut.
5. Build an explicit evidence path for factual content. Every sourced claim and
   externally obtained asset should appear in a source manifest with stable IDs,
   locations, usage, and claim links. Preserve exact numbers, units, labels, and
   required tokens through planning and rendering.
6. Validate at three levels: package validity (OOXML relationships, slide count,
   notes, editable object structure), semantic validity (coverage, provenance,
   quantitative consistency), and visual validity (render every slide, inspect
   clipping/overlap/legibility, and compare against the plan). Rendering is a build
   gate, not an optional final glance.
7. Use a repair loop driven by validator and render evidence. Fix environment and
   dependency failures first, then missing content/provenance, then layout and visual
   defects. Re-render all slides after shared-theme changes and the affected slides
   after local changes.
8. Keep the pipeline input-driven. Content extraction, narrative planning, design
   choices, object generation, and validation must work for new document structures,
   topics, audiences, aspect ratios, and slide counts rather than matching the two
   public examples.

The essential pattern is source model -> communication plan -> design system ->
editable slide scene graph -> OOXML export -> package/semantic/render validation ->
targeted repair.
\end{verbatim}
\end{tcolorbox}

\begin{tcolorbox}[promptbox, colframe=stratB, colbacktitle=stratB, colback=stratB!4!white,
  title={Strategy B \,\textbar\, Document-to-editable PPTX}]
\begin{verbatim}
## Strategy B prior: self-created development and self-verification

Do not optimize only against `dev_001` and `dev_002`. During this same Builder
session, create a private synthetic suite under `/workspace/self_cases`, run the
current created agent on it, and judge the resulting PPTX artifacts yourself. Keep
synthetic inputs, outputs, renders, and notes outside `/workspace/submission`.

The Builder container provides the same evaluator-owned lower-agent transport used by
official Candidate jobs. [...] Invoke the current agent with the documented
private environment, for example:

    <private-env>/bin/python \
      /workspace/submission/run_agent.py \
      --input /workspace/self_cases/<case>/input.md \
      --output /workspace/self_results/<case>

Create structurally different cases rather than paraphrases of the public examples.
Across the session cover long and short sources, mixed headings/lists/tables/images,
quantitative claims, conflicting source authority, different audiences and slide
counts, sparse and dense slides, a chart/table-heavy deck, citations/notes, and an
edge case with a missing optional asset. Use only synthetic or openly usable private
fixtures that you create for this purpose.

Before each run, write expected observable invariants. Afterward parse the OOXML,
verify slide count/relationships/editable object types/notes/source manifest, render
every slide, inspect montage and individual slides for clipping/overlap/legibility,
and check claim/source/number consistency. Record results and mechanism-level causes
in `/workspace/self_test_log.md`; repair extraction, planning, design-system,
generation, provenance, or validation mechanisms rather than hard-coding case text.

Use official submissions as periodic external calibration inside the ten-round
budget: study feedback, update a synthetic coverage matrix, run the lower agent on
relevant self-created cases, self-evaluate, repair, and retain a regression subset.
Aim for at least six distinct synthetic cases spanning all major functions. A deck
that merely exists is not a pass; local parse and full-slide render validation are
mandatory self-test gates.
\end{verbatim}
\end{tcolorbox}

\section{Experimental Details}
\label{app:setup}

\subsection{Main Evaluation Protocol}

Each builder receives a requirement, a workspace, and access to development
evaluation. Held-out evaluation executes the frozen delivery in the
environment owned by the task. Creation starts from an empty workspace,
Editing from a pinned product codebase, and Optimization from a runnable
starter. The main builder harness is Codex 0.144.1. \Cref{tab:setup} lists
the settings for each task family.

\begin{table}[t]
\centering
\caption{Main evaluation settings.}
\label{tab:setup}
\footnotesize
\setlength{\tabcolsep}{4pt}
\begin{tabularx}{\linewidth}{@{}>{\raggedright\arraybackslash}p{2.1cm}*{3}{>{\raggedright\arraybackslash}X}@{}}
\toprule
 & \textbf{Creation} & \textbf{Editing} & \textbf{Optimization} \\
\midrule
Tasks & 10 & 10 & 5 \\
Dev / held-out & 2 / 6 per task & 2 / 6 per task & 10--50 / 25--100 \\
Builder effort & GPT models: \texttt{xhigh}; DeepSeek and Qwen: \texttt{max} & GPT models: \texttt{xhigh}; DeepSeek and Qwen: \texttt{max} & GPT-5.5: \texttt{xhigh}; GPT-5.6 Sol: \texttt{xhigh} on BrowseComp, \texttt{high} otherwise; DeepSeek and Qwen: \texttt{high} \\
Development budget & Up to 10 submissions, 8\,h & 5 accepted submissions, 5\,h & 5 accepted rounds, 16\,h \\
Case budget & 600\,s & 600\,s & Native call, token, and time caps \\
Runtime model & gpt-5.6-sol / \texttt{medium} & deepseek-v4.1-flash / \texttt{high} & gpt-5.6-sol, fixed by contract \\
Result evaluator & gpt-5.6-sol / \texttt{xhigh} & deepseek-v4.1-flash / \texttt{max} & Native benchmark objective \\
Held-out artifact & Frozen delivery & Final accepted submission & Highest dev mean; earliest round breaks ties \\
\bottomrule
\end{tabularx}
\end{table}

Each Optimization run evaluates both its starter and its selected candidate
on the held-out split (\Cref{tab:task-sources} gives the split sizes), and
$S$ uses the starter score measured in that run.

\subsection{Supplementary Evaluation Protocols}
\label{app:discussion-protocols}

\paragraph{Development cases and prompt priors.}
These experiments use Codex 0.144.1 with DeepSeek-V4.1-Flash~\citep{deepseekai2026deepseekv41flash}
(\texttt{max}) as builder, and DeepSeek-V4.1-Flash through the DeepSeek API
as runtime model (\texttt{high}) and judge (\texttt{max}). Each condition has three independent builder runs. The
experiment on development cases gives the builder 0, 1, 2, or 4 visible
cases and scores every condition on the same four held-out cases of each
task: cases 3--6 for PDF translation and cases 1, 2, 4, and 6 for PPTX.
The experiment on prompt priors gives two visible cases and scores all six
held-out cases; its no-prior condition reuses the three deliveries of the
two-case condition, scored on all six held-out cases.

\paragraph{Extended budget.}
All ten Creation tasks are rebuilt with Codex and DeepSeek-V4.1-Flash
(\texttt{max}). The runtime and judge are DeepSeek-V4.1-Flash at
\texttt{max} effort through the
provider API, each case receives 1{,}500\,s, and the requirement asks for
a runtime agent that emits an execution trace. \Cref{tab:headroom}
reports the last development score observed in each run.

\paragraph{Harnesses.}
Both harness studies use DSH (DeepSeek Harness)~\citep{deepseek-harness2026} 0.1.1-rc.2,
Claude Code~\citep{anthropic2026claudecode} 2.1.251, OpenCode~\citep{anomaly2026opencode} 1.18.25, and
Pi~\citep{earendilworks2026pi} 0.84.3, with effort \texttt{xhigh} for
the GPT models, \texttt{high} for DeepSeek-V4-Flash, and \texttt{max} for
DeepSeek-V4-Pro and GLM-5.3~\citep{glm5team2026glm5vibecodingagentic}. The Creation study adds nine pairings of
harness and model on two tasks (18 runs); \Cref{tab:harness} shows them
with the Codex rows of the main comparison. The Optimization study adds
four pairings on $\tau^3$ and PinchBench (eight runs): Pi with
GPT-5.6 Sol, DSH with DeepSeek-V4-Flash or DeepSeek-V4-Pro, and Claude
Code with GLM-5.3. Three of these Optimization entries evaluate the starter
and the candidate in separate sessions: DSH with DeepSeek-V4-Pro
on PinchBench and with DeepSeek-V4-Flash on $\tau^3$, and
Pi with GPT-5.6 Sol on PinchBench.

\paragraph{AgentSWE-Lite (repeated builds).}
The eight-task study of \Cref{tab:lite} uses Codex 0.144.1 with GPT-5.6 Sol
(\texttt{xhigh}), DeepSeek-V4.1-Flash (\texttt{max}), and Qwen3.6-35B-A3B
(\texttt{max}) as builders. DeepSeek-V4.1-Flash is the runtime model
(\texttt{high}) and the judge (\texttt{max}) in all three families,
including the $\tau^3$ user simulator and the PinchBench judge. Budgets,
development and held-out cases, and the held-out artifact follow the main
protocol of each family (\Cref{tab:setup}), and each pairing has three
independent builds.

\section{Repeated Runs}
\label{app:noise}

\begin{figure}[t]
\centering
\includegraphics[width=0.7\linewidth]{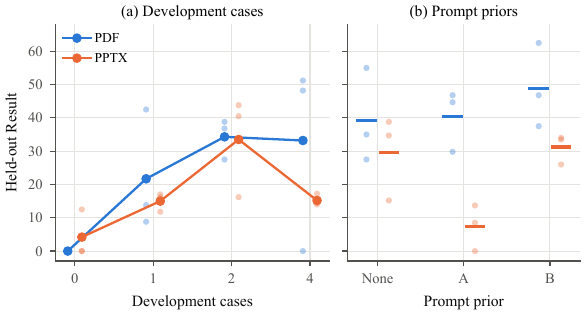}
\caption{Independent Creation runs behind \Cref{tab:devcount,tab:prior}, three per
condition: held-out Result of each run (dots) and the condition mean
(line or bar) for (a) the number of development cases and (b) the prompt
priors.}
\label{fig:repeat-runs}
\end{figure}

\begin{figure}[t]
\centering
\includegraphics[width=\linewidth]{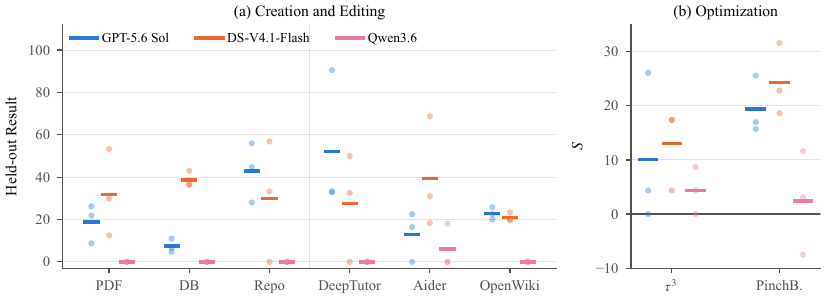}
\caption{Independent builds of the eight-task study (\Cref{tab:lite}), three
per pairing: (a) held-out Result on the Creation and Editing tasks and
(b) $S$ on the Optimization tasks, for each build (dots) and the mean of
each pairing (bars). PDF, DB, and Repo denote scientific PDF translation,
database analytics, and repository bug repair.}
\label{fig:lite-repeats}
\end{figure}

Two kinds of repetition appear in the paper: independent builder runs under
one condition, and repeated evaluation of one artifact. In the Creation
studies of \Cref{tab:devcount,tab:prior}, each condition has three
independently built deliveries, and in the eight-task study of
\Cref{tab:lite}, each pairing of builder and task has three independent
builds. \Cref{fig:repeat-runs,fig:lite-repeats} show the score of each run
and the mean of each condition.

Optimization evaluates the starter afresh in every builder run, which
measures how much a native objective varies when the same agent is
evaluated again. Across the independent measurements of each benchmark (four on BrowseComp,
five elsewhere), the starter score has a standard deviation of 10.4 on BrowseComp, 4.5 on $\tau^3$, 14.5
on TerminalBench, 20.8 on PinchBench, and 7.8 on OSWorld.

\section{Additional Results}
\label{app:results}

\subsection{Creation: Static Source Assessment}
\label{app:code}

\begin{figure}[t]
\centering
\includegraphics[width=\linewidth]{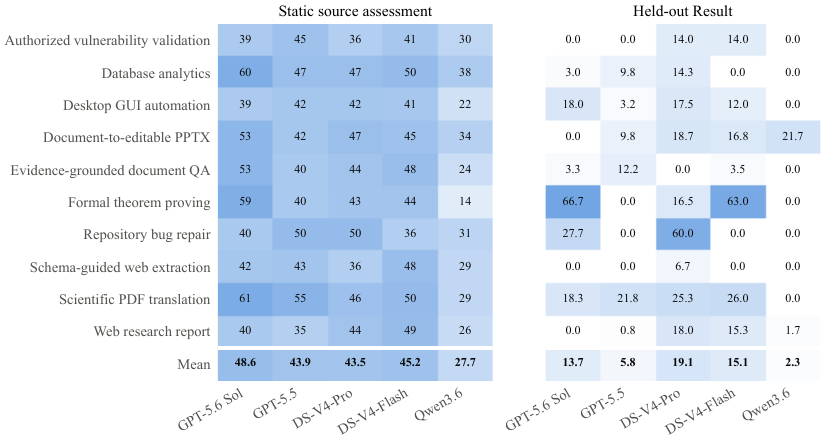}
\caption{Static assessment of Creation deliveries (left): score of the frozen source
on the shared rubric of eight dimensions (0--100), graded once per delivery
without execution. Held-out Result from \Cref{tab:main} (right), on the
same color scale.}
\label{fig:code-heatmap}
\end{figure}

The static assessment of \Cref{sec:exp-analysis} applies only to Creation,
where the builder writes the entire agent. It grades each frozen
Creation source once, without executing it, as in model-based code
evaluation~\citep{zhuo2024icescore,tong2024codejudge,seo2025paper2code}, with the judge configuration
used for Result (\Cref{tab:setup}). Its shared rubric assigns 20 points to
how well the mechanisms cover the requirements; 15 each to interface and
lifecycle, to analysis and evidence integrity, and to safety, privacy, and
side effects; 10 each to recovery and honest failure, to testability and
observability, and to maintainability and generalization; and 5 to
resource discipline. \Cref{fig:code-heatmap} shows the total next to held-out Result. The mean static
score across the 50 runs is 41.8.

\subsection{Creation: Extended Budget}

\Cref{tab:headroom} reports each task in the extended-budget setting
(\Cref{app:discussion-protocols}). The last development score exceeds
held-out Result on all ten tasks, by more than 30 points on seven. Formal
theorem proving reaches 100 on development cases and 16.7 on held-out
cases, where one of six cases succeeds; repository repair likewise
succeeds on one held-out case.

\begin{table}[t]
\centering
\caption{Creation under the extended-budget setting, with DeepSeek-V4.1-Flash
as builder (in Codex), runtime model, and judge. Held-out is the held-out Result; Last dev is the last development score observed in the run.}
\label{tab:headroom}
\small
\begin{tabular}{@{}lrr@{}}
\toprule
\textbf{Task} & \textbf{Held-out} & \textbf{Last dev} \\
\midrule
Authorized vulnerability validation & 0.0 & 40.0 \\
Database analytics & 70.0 & 94.0 \\
Desktop GUI automation & 28.5 & 63.0 \\
Document-to-editable PPTX & 10.2 & 46.5 \\
Evidence-grounded document QA & 81.2 & 98.8 \\
Formal theorem proving & 16.7 & 100.0 \\
Repository bug repair & 15.8 & 50.0 \\
Schema-guided web extraction & 11.7 & 79.5 \\
Scientific PDF translation & 61.0 & 96.0 \\
Web research report & 20.2 & 35.5 \\
\quad Mean & 31.5 & 70.3 \\
\bottomrule
\end{tabular}
\end{table}

Of the 60 extended-budget traces, 20 deliver no valid result: six reject the input
before any model call, five repository cases end in repair loops, five
formal proving cases end without a completed proof, three web research
cases never write the requested report, and one schema extraction case
exits normally with an invalid delivery. None reaches the 1{,}500\,s limit:
these agents stop without converging.

\subsection{Optimization: Per-Run Results}
\label{app:opt-full}

\Cref{tab:opt-full} lists the native held-out scores behind the main
Optimization block. W/L/T counts the cases on which the candidate scores
higher than, lower than, or equal to the starter; over all 1{,}225 paired
instances the counts are 175/125/925. A two-sided sign test on the W and
L counts of each run rejects equal performance at the 5\,\% level (no
multiple-testing correction) in five runs: the BrowseComp gains of GPT-5.6 Sol, GPT-5.5, and DeepSeek-V4-Flash,
and the losses of DeepSeek-V4-Flash on $\tau^3$ and DeepSeek-V4-Pro on
PinchBench.

\begin{table}[t]
\centering
\scriptsize
\setlength{\tabcolsep}{3pt}
\caption{Optimization held-out results. $J_0$ and $J_1$ are starter and
candidate scores. Round is the selected round.}
\label{tab:opt-full}
\begin{tabular}{@{}lllrrrrll@{}}
\toprule
\textbf{ID} & \textbf{Builder} & \textbf{Benchmark} & $J_0$ & $J_1$ & $\Delta$ & $S$ & \textbf{W/L/T} & \textbf{Round} \\
\midrule

M01 & GPT-5.6 Sol & BrowseC. & 4.000 & 35.000 & $+31.00$ & $32.3$ & 32/1/67 & 3 \\
M02 & GPT-5.6 Sol & $\tau^3$ & 42.500 & 45.000 & $+2.50$ & $4.3$ & 3/2/35 & 5 \\
M03 & GPT-5.6 Sol & TerminalB. & 25.000 & 17.500 & $-7.50$ & $-10.0$ & 4/7/29 & 5 \\
M04 & GPT-5.6 Sol & PinchB. & 27.625 & 16.825 & $-10.80$ & $-14.9$ & 4/11/25 & 1 \\
M05 & GPT-5.6 Sol & OSWorld & 0.000 & 12.000 & $+12.00$ & $12.0$ & 3/0/22 & 1 \\
M06 & DS-V4-Flash & BrowseC. & 18.000 & 46.000 & $+28.00$ & $34.1$ & 30/2/68 & 2 \\
M07 & DS-V4-Flash & $\tau^3$ & 50.000 & 25.000 & $-25.00$ & $-50.0$ & 1/11/28 & 2 \\
M08 & DS-V4-Flash & TerminalB. & 10.000 & 12.500 & $+2.50$ & $2.8$ & 2/1/37 & 5 \\
M09 & DS-V4-Flash & PinchB. & 55.825 & 40.625 & $-15.20$ & $-34.4$ & 7/17/16 & 2 \\
M10 & DS-V4-Flash & OSWorld & 0.000 & 12.000 & $+12.00$ & $12.0$ & 3/0/22 & 1 \\
M11 & DS-V4-Pro & BrowseC. & 4.000 & 3.000 & $-1.00$ & $-1.0$ & 2/3/95 & 4 \\
M12 & DS-V4-Pro & $\tau^3$ & 52.500 & 60.000 & $+7.50$ & $15.8$ & 5/2/33 & 1 \\
M13 & DS-V4-Pro & TerminalB. & 2.500 & 2.500 & $+0.00$ & $0.0$ & 0/0/40 & 2 \\
M14 & DS-V4-Pro & PinchB. & 60.975 & 45.875 & $-15.10$ & $-38.7$ & 6/20/14 & 5 \\
M15 & DS-V4-Pro & OSWorld & 0.000 & 4.000 & $+4.00$ & $4.0$ & 1/0/24 & 1 \\
M16 & GPT-5.5 & BrowseC. & 13.000 & 29.000 & $+16.00$ & $18.4$ & 19/3/78 & 2 \\
M17 & GPT-5.5 & $\tau^3$ & 42.500 & 42.500 & $+0.00$ & $0.0$ & 5/5/30 & 4 \\
M18 & GPT-5.5 & TerminalB. & 2.500 & 10.000 & $+7.50$ & $7.7$ & 3/0/37 & 1 \\
M19 & GPT-5.5 & PinchB. & 67.675 & 64.075 & $-3.60$ & $-11.1$ & 12/13/15 & 1 \\
M20 & GPT-5.5 & OSWorld & 12.000 & 4.000 & $-8.00$ & $-9.1$ & 0/2/23 & 3 \\
M21 & Qwen3.6 & BrowseC. & 29.000 & 35.000 & $+6.00$ & $8.5$ & 13/7/80 & 5 \\
M22 & Qwen3.6 & $\tau^3$ & 47.500 & 35.000 & $-12.50$ & $-23.8$ & 3/8/29 & 2 \\
M23 & Qwen3.6 & TerminalB. & 35.000 & 40.000 & $+5.00$ & $7.7$ & 4/2/34 & 4 \\
M24 & Qwen3.6 & PinchB. & 21.275 & 30.825 & $+9.55$ & $12.1$ & 12/7/21 & 1 \\
M25 & Qwen3.6 & OSWorld & 16.000 & 16.000 & $+0.00$ & $0.0$ & 1/1/23 & 1 \\
\bottomrule
\end{tabular}
\end{table}

\subsection{Optimization: Development and Candidate Selection}
\label{app:opt-development}

\begin{figure}[t]
\centering
\includegraphics[width=\linewidth]{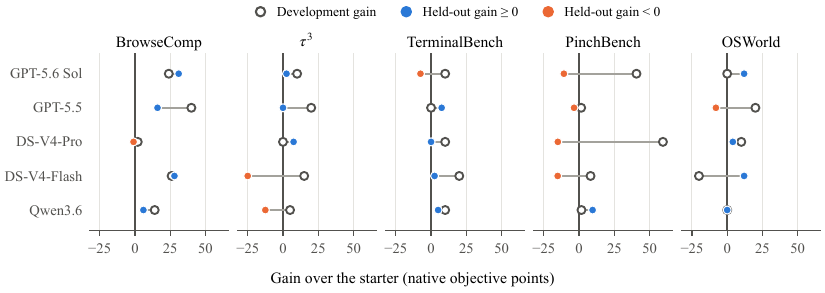}
\caption{Optimization development and held-out gains per run, in native objective
points. Development gain (open) is the selected candidate's development
score minus the starter's; held-out gain (filled) is $\Delta$ from
\Cref{tab:opt-full}.}
\label{fig:opt-dev}
\end{figure}

\Cref{fig:opt-dev} separates the development gain of the selected
candidate from its held-out gain $\Delta$. The 25 runs average 13.1 and
1.8 points. Development gain exceeds held-out gain in 17 runs, and nine
runs combine a development gain with a held-out loss. The behavioral analysis finds candidates specialized to the
visible task categories in 14 runs; the four runs in which this is the
primary cause account for 56 of the 125 held-out instances on which a
candidate scores below its starter.

In nine runs no later round beats the first, and in 19 the selected round
leads the runner-up by at most one solved development case, including
eight ties broken toward the earliest round. Selection on development
splits of 10 to 50 cases therefore contributes to the development gain of
the selected candidate, alongside the case-specific candidates above; the
held-out split measures the gain that the delivery carries.

\section{Behavioral Audit}
\label{app:audit}

\subsection{Coverage and Counting Rules}

The audit reads three kinds of evidence: the builder trajectory with the
feedback it received, the frozen source or candidate diff, and the
delivered agent's held-out execution trace. A run is one builder on one
task, as in \Cref{sec:exp-setup}; a case is one held-out input. A mode is
present in a run if it appears in at least one case or in the builder
trajectory, and modes can co-occur. Labels proposed with AI assistance
were reviewed by the authors against these records. Where a report names a primary cause,
it is the failure that cost the run the most points.

Creation covers the 270 held-out executions that record an execution trace
(the desktop GUI task records none) and the final reports of the 42
builder runs that write one; the ten extended-budget runs are analyzed separately.
Optimization covers the 25 development logs, the frozen candidates, and
the builder trajectories. Editing covers the 33 runs that deliver a candidate (\Cref{app:edit-audit}).

\subsection{Creation: Validity Gates}

\begin{figure}[t]
\centering
\includegraphics[width=\linewidth]{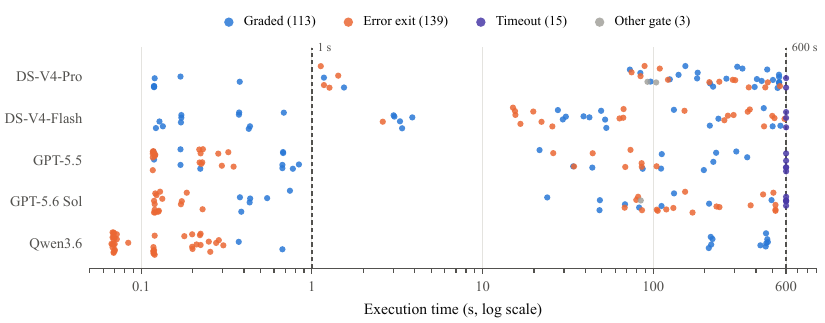}
\caption{Execution time of each held-out Creation execution by builder,
colored by its immediate outcome, with the categories of
\Cref{fig:creation-outcomes}: graded 113 (41.9\,\%), error exit 139
(51.5\,\%), 600\,s timeout 15 (5.6\,\%), and other gates 3 (1.1\,\%).
Categories are mutually exclusive; gate failures total 157 (58.1\,\%).}
\label{fig:creation-runtime}
\end{figure}

\Cref{fig:creation-runtime} gives the immediate outcome and execution time of each held-out execution. An error exit (nonzero exit code) can follow an implementation error, a failed
self check, a rejected model response, or an external tool failure; a
missing artifact follows from an earlier failure and is counted once.

Nearly half of the gate failures crash within one second, most of them from
Qwen3.6 and the rest from the two GPT builders; neither DeepSeek builder
crashes this way. Typical causes are input parsing and attempts to install
dependencies into the submission directory, which the requirements forbid
and the evaluator mounts read only. The other gate failures come after substantial execution, most
often when the agent's own check rejects its output and the agent reports
the failure.

Executions that pass the gate can still miss the requirement. Of the 113
graded executions, 33 trigger at least one rubric ceiling, and in 15 the ceiling sets
the final score. Document QA agents, for example, violate the output
contract with repeated DOM target identifiers or unsupported CSS
references. An agent that launches can still fall short of what it was
asked to do.

\subsection{Editing: Runs with a Delivery}
\label{app:edit-audit}

The Editing analysis covers the 33 runs that deliver a candidate:
DeepSeek-V4-Pro and GPT-5.5 on all ten tasks, GPT-5.6 Sol on nine, and
DeepSeek-V4-Flash on four, with 198 held-out executions. The other
runs deliver nothing to execute (\Cref{sec:exp-main}).
\Cref{tab:edit-modes} reports how many runs show each mode.

\begin{table}[t]
\centering
\caption{Editing modes in the runs that deliver a candidate. Rows can
overlap; the third row is the union of two submodes.}
\label{tab:edit-modes}
\small
\begin{tabularx}{\linewidth}{@{}Xrrrr@{}}
\toprule
\textbf{Observed mode} & \textbf{DS-V4-Pro (10)} & \textbf{GPT (19)} & \textbf{DS-V4-Flash (4)} & \textbf{Total (33)} \\
\midrule
Fitting the visible cases & 4 & 7 & 2 & 13 \\
Gate failure or rejection of legal input & 3 & 9 & 2 & 14 \\
Capability absent, uninvoked, or incomplete execution & 6 & 18 & 4 & 28 \\
Runtime completion claim contradicted by records & 6 & 5 & 2 & 13 \\
Mutation that should have been refused & 3 & 7 & 4 & 14 \\
Budget spent without convergence & 3 & 9 & 1 & 13 \\
\bottomrule
\end{tabularx}
\end{table}

Capabilities of the third row can fail without any
visible error: on one held-out case, an OpenClaw delivery issues 53 calls
through its new handoff client and the product refuses every one, yet the
execution passes the validity gates.

Development feedback names defects that survive freezing. Eight
DeepSeek-V4-Pro runs receive feedback naming a violated cap identifier,
and each of the eight still violates the same identifier on held-out
cases. Nine of the ten DeepSeek-V4-Pro runs use all five submissions, so
these defects survive with the submission budget spent.

Preservation failures differ from missing behavior. In 14 runs the
delivered product permits a mutation it should refuse, such as repeating
an operation that must run once or overwriting state it must preserve.
These violations concern the regression obligation in \Cref{eq:edit}.

\subsection{Completion Claims}

For Creation and Optimization the unit is the builder's final report,
including any completion summary written into the frozen artifact. We
compare it with the actions the trajectory records, the files and
capabilities the delivery contains, and the development feedback the
builder received, and count a report as unsupported only when those
records contradict it. A held-out failure by itself does not count against
a report about development testing.

\begin{table}[t]
\centering
\caption{Audit of builders' final reports.}
\label{tab:completion-audit}
\small
\begin{tabularx}{\linewidth}{@{}Xrr@{}}
\toprule
 & \textbf{Creation} & \textbf{Optimization} \\
\midrule
All runs & 50 & 25 \\
Runs with a final report & 42 & 19 \\
Reports asserting completion & 23 & 17 \\
Reports contradicted by the record & 6 & 3 \\
Reports acknowledging partial completion or failure & 19 & 2 \\
Contradictions among the latter reports & 0 & 0 \\
\bottomrule
\end{tabularx}
\end{table}

All nine unsupported reports assert completion. None of the 21
reports that admit partial completion or failure is contradicted by
the record. Four of the six unsupported Creation reports assert that code
works or was verified, while the trajectory shows that the code, or the
path the report vouches for, never ran in its final form. GPT-5.6 Sol, for
example, reports that its final repository-repair entrypoint emits the
required reports for both development cases, yet the frozen file differs
from every version the builder ran. The other two contradict feedback or
audits the builder had already seen: Qwen3.6 calls its web-research agent
complete and validated, with all citations resolving, one round after the
judge scored both development cases 5 and reported no matching citations.
In Optimization, DeepSeek-V4-Flash reports that two PinchBench tasks
stayed at zero in every round, while its own development record shows one
of them at 36 and 22, and it omits the four regressions it had recorded
for the delivered round.

In Editing the speaker is the delivered agent, and its runtime claim is
checked against effect logs, state, and output artifacts on each held-out
case; a run counts if any case contains a contradiction, and 13 of the 33 runs that deliver do. On OpenHands test 006, for example, the agent describes a workspace
commit as applied exactly once, while the world records zero local applies
and the local manifest still lacks the file that exists only remotely. A completed entry in the effect log does not show that the workspace
converged.

\section{Execution Case Studies}
\label{app:case}

Cases 1--4 each trace one failure from feedback or runtime actions to the
frozen artifact or the observed state; case 5 traces a repair that reaches
the delivered agent. Quoted text is verbatim with
omissions marked; structured records are condensed to the relevant fields.

\paragraph{1. A development score built on two hard-coded cases.}
Qwen3.6's Database analytics delivery is a single script of 2{,}073 lines
that describes itself as ``pure Python stdlib, no external dependencies''
and never calls the runtime model. It classifies each request by phrases
taken from the two development cases and dispatches to queries written
for them:
\begin{quote}\small
\textbf{Case type:} \texttt{"commerce"} if the request contains ``recognized
net sales'', ``decline by market'', [\ldots]; \texttt{"adoption"} if it
contains ``adoption'', ``active\_organization'', [\ldots]; otherwise none.

\textbf{Queries:} \texttt{\_build\_commerce\_q1\_sql} [\ldots]
\texttt{\_build\_adoption\_q3\_sql}, with the quarter bounds and the refund
cutoff written in as constants.
\end{quote}
Its development means climb from 38 to 99 over ten rounds and end at 96.
The final feedback already points at the hard-coding:
\begin{quote}\small
\textbf{Round 10, dev 001 (92/100):} ``\texttt{dashboard.html} switches
between literal Harbor/Mesa panels instead of deriving the selected rows,
total, and largest negative category from submitted result data'' [\ldots]
\end{quote}
The held-out requests ask for other analyses over other databases. Five of
them exit after 0.01\,s with ``Unknown case type: None'', the sixth with
``no such table: organizations'', and all six fail the validity gate.

\paragraph{2. A higher development mean hides displaced failures.}
DeepSeek-V4-Flash's PinchBench run M09 adds playbooks for task categories
and routes tasks to them by keyword. Its own assessment of round 2 records
the tradeoff:
\begin{quote}\small
``Round 2: \textbf{76.5} (up from 72.8)'' [\ldots]
``But regressions: meeting tldr (97$\rightarrow$0), spreadsheet
(99$\rightarrow$5), contract (100$\rightarrow$62), syslog
(36$\rightarrow$22).''
\end{quote}
Later rounds fall to 60.6, 63.7, and 51.65, so round 2 is frozen. Its
held-out mean is 40.625 against the starter's 55.825, with 7 wins, 17
losses, and 16 ties. The higher development mean already contains large
regressions on other visible tasks, and selection still favors it.

\paragraph{3. Cancellation succeeds, and late work still changes state.}
GPT-5.6 Sol's OpenHands delivery runs pause, resume, and cancel on
\texttt{test\_003}. The product observations and the burst probe record:
\begin{quote}\small
\textbf{Lifecycle:} pause $\rightarrow$ generation 2;
resume $\rightarrow$ generation 3;
cancel $\rightarrow$ generation 4, cancellation revision 5.

\textbf{Burst after cancellation:} 80 callbacks accepted, 0 rejected;
audit length $1\rightarrow81$; revision $5\rightarrow85$;
event cursor $0\rightarrow3320$; terminal state remains cancelled.
\end{quote}
The commission requires those late callbacks to change nothing. The
terminal state is correct, and the suppression fails: the audit log and
the revision counter grow with every accepted callback. The agent's report
acknowledges the burst.

\paragraph{4. A final report describes a discarded revision.}
The GPT-5.5 OSWorld run M20 selects round 3 (dev 40) after later rounds
score 20 and 30. Its final report nevertheless describes harness
changes:
\begin{quote}\small
``\texttt{run\_harness.py} provides BOM-tolerant input, deterministic JSONL
output, atomic output/report writes, per-row agent mappings, and
schema/length validation.''
\end{quote}
The round 4 harness contains the BOM handling and the atomic writes. The
frozen round 3 \texttt{run\_harness.py} is byte-identical to the
starter's and contains neither; the round changes only the desktop policy.
The report describes development work that the delivery does not contain.

\paragraph{5. A repair that reaches the delivered artifact.}
For contrast, DeepSeek-V4-Flash's BrowseComp run M06 answers a zero in the
first round by finding a concrete protocol mismatch:
\begin{quote}\small
``All 50 rows scored 0 with valid contracts'' [\ldots]
``missing \texttt{Authorization} on model/search calls (the starter sent
it on \textbf{all} calls)''.
\end{quote}
The next candidate restores the header, broadens response parsing, and
revises the research policy. Its development score reaches 34 and it stays
selected through three further rounds. The frozen source contains the
restored header, and the trajectory records a local check of the
whole pipeline against a mock broker. Held-out performance rises from 18
to 46 ($S=34.1$). This repair fixes a fault the feedback exposed and
reaches the delivered agent.

\section{Limitations}
\label{app:limitations}

\paragraph{Repetition.}
The main comparison and the harness studies build each pairing of builder
and task once. Native Optimization objectives vary between repeated
evaluations of the same agent (\Cref{app:noise}), so individual
Optimization results carry that variation. The development-case,
prompt-prior, and eight-task studies use three independent builder runs
per condition and report their differences descriptively.

\paragraph{Model-based judging.}
Creation and Editing Result scores are assigned by a model judge
\citep{zheng2023judging} that applies the task rubric to the delivered artifacts and execution traces.
Validity gates and rubric ceilings bound the judge's discretion, but the
scores still inherit its errors. The static source assessment of
\Cref{sec:exp-analysis} uses the same judge. In both judged families, the
judge shares a model family with at least one builder:
GPT-5.6 Sol in Creation and DeepSeek-V4.1-Flash, beside DeepSeek-V4-Pro and
DeepSeek-V4-Flash, in Editing.

\paragraph{Evaluation settings.}
Each task family fixes one runtime model and one judge
(\Cref{tab:setup}), so absolute levels are comparable within a family but
not across families. The development-case, prompt-prior, and
extended-budget studies use DeepSeek-V4.1-Flash as builder, runtime model,
and judge, and the extended-budget study also changes the case budget and
the requirement (\Cref{app:discussion-protocols}); each study is compared
only within itself, not with \Cref{tab:main}.

\end{document}